\documentclass[preprint]{aastex}
\shorttitle{}
\shortauthors{Lee}
\begin{document}
\title{On the Intrinsic Alignments of the Late-Type Spiral Galaxies from the 
Sloan Digital Sky Survey Data Release 7}
\author{Jounghun Lee}
\affil{Astronomy Program, Department of Physics and Astronomy, FPRD, Seoul National 
University, Seoul 151-747, Korea }
\email{jounghun@astro.snu.ac.kr}
\begin{abstract}
A robust detection of the tidally induced intrinsic alignments of the late-type spiral 
galaxies with high statistical significance is reported. From the spectroscopic galaxy sample 
of SDSS DR7 compiled by Huertas-Company et al. which lists each galaxy's probabilities 
of being in five Hubble types, $P({\rm E}),\ P({\rm Ell}), \ P({\rm S0}),\ P({\rm Sab}),\ 
P({\rm Scd})$, we select the nearby large late-type spiral galaxies which have redshifts 
of $0\le z\le 0.02$, probabilities of $P({\rm Scd})\ge 0.5$ and angular sizes of $D\ge 7.92$ arcsec. 
The spin axes of the selected nearby large late-type spiral galaxies are determined up to 
the two-fold ambiguity with the help of the circular thin-disk approximation and their spatial correlations 
are measured as a function of the separation distance $r$. A clear signal of the intrinsic correlation as high 
as $3.4\sigma$ and $2.4\sigma$ is found at the separation distance of $r\approx 1\,h^{-1}$Mpc and 
$r\approx 2\,h^{-1}$Mpc, respectively. The comparison of this observational results with the analytic model 
based on the tidal torque theory reveals that the spin correlation function for the late-type spiral galaxies follow 
the quadratic scaling of the linear density correlation and that the intrinsic correlations of the galaxy spin axes 
are stronger than that of the underlying dark halos. We investigate a local density dependence of the galaxy spin 
correlations and found that the correlations are stronger for the galaxies located in dense regions having more 
than $10$ neighbors within $2\,h^{-1}$Mpc. We also attempt to measure a luminosity dependence of the galaxy spin 
correlations, but find that it is impossible with our magnitude-split samples to disentangle a luminosity from a redshift 
dependence. We provide the physical explanations  for these observational results and also discuss the effects of possible 
residual systematics on the results.
\end{abstract}
\keywords{cosmology:theory --- methods:statistical --- large-scale structure of 
universe}

\section{INTRODUCTION}

The building blocks of the large-scale structure in the universe are the galactic dark 
halos that are believed to underlie the luminous galaxies. Although the galactic dark 
halos are not directly observable, their intrinsic properties can be inferred and 
estimated from the galaxy observables such as luminosity, color, morphology and spatial 
correlations. For instance, how luminous a given galaxy is related to how massive its  
underlying dark halo is. The colors and morphologies of the observed galaxies reflect 
the formation epochs and assembly history of their galactic halos. Finding the hidden 
relations between the galaxy observables and intrinsic properties of the underlying 
dark halos is one of the most important tasks in the field of the large-scale 
structure. 

The directions of the galaxy angular momentum (i.e., the galaxy spin axes) are one of 
those galaxy observables that are believed to have a direct connection with the 
intrinsic properties of the underlying dark halos. In the standard linear tidal torque 
theory \citep{pee69,dor70,whi84,dub92,CT96}, the tidal torques from the surrounding 
matter originate the angular momentum of the proto-galactic halos in the linear regime 
\citep[see][for a recent review]{sch09}. In consequence, the directions of the 
angular momentum of the proto-galactic halos come to be aligned with the intermediate 
principal axes of the local tidal tensors \citep{LP00,LP01,LP02,por-etal02}. 
As the tidal fields are spatially correlated, the correlations between the spin axes of the 
proto-galactic halos and the principal axes of the local tidal tensors would induce the 
spatial correlations of the spin axes of the proto-galactic halos. A detailed analysis have 
shown that the spatial correlations of the spin axes of the galactic halos would follow a 
quadratic scaling with the two point correlation function of the linear density field 
provided that the linear tidal torque theory is valid 
\citep{pen-etal00,sug-etal00,LP01,cri-etal01}.
 
A critical question is whether or not the tidally induced initial correlations between the spin 
axes of the galactic halos have been retained to a detectable level and how strong the retained 
correlations are. If the intrinsic correlations of the spiral galaxies are detected and if we can model 
them well with the tidal torque theory, it would make it plausible  to reconstruct the initial tidal 
fields from the observable spin fields as proposed by \citet{LP00}. Furthermore, it would have a 
direct impact on the weak lensing community where the intrinsic alignments of the spiral (or blue) 
galaxies are often assumed to be zero \citep[e.g,][]{joa-etal10,man-etal10}.

Plenty of observational efforts have so far been made to address this issue. 
\citet{pen-etal00} measured the spatial correlations of the spin axes of the nearby 
spiral galaxies from the Tully catalog and reported a tentative detection of $2\sigma$ 
signals. They claimed that the observed trend of the galaxy spin correlations is consistent 
with the quadratic scaling of the linear density two-point correlations. \citet{LP02} searched 
directly for the correlations between the spin axes of the nearby spiral galaxies and the 
intermediate principal axes of the tidal fields using the data from the Point Source Galaxy 
catalog \citep{pscz}. 

\citet{brown-etal02} demonstrated that the intrinsic shape correlations of the elliptical 
galaxies from the SuperCOSMOS survey \citep{supercosmos} are consistent with the quadratic 
scaling with the linear density correlations under the assumption that the projected minor axes of the 
observed elliptical galaxies are orthogonal to the projected spin axes of their underlying halos. 
\citet{nav-etal04} noted that the distribution of the inclination angles of the nearby disk galaxies 
relative to the Supergalactic plane is consistent with the prediction of the linear tidal torque theory.  
\citet{tru-etal06} claimed that the observed alignments of the galaxies in the 
vicinity of large voids are well explained by the tidally induced intrinsic correlations. 
\citet{LE07} detected a $2\sigma$ signal of the correlations between the spin axes of the nearby 
spiral galaxies from the Tully catalog and the intermediate principal axes of the local tidal field 
constructed from the 2MASS redshift survey \citep[][and references therein]{2mass}.

\citet{LP07} measured the two dimensional projected spin correlations of the blue galaxies 
from the Sloan survey \citep{sdss} and found a $3\sigma$ signal at one single distance bin 
of $r\sim 1\,h^{-1}$Mpc. \citet{paz-etal08} found the correlations between the galaxy angular 
momentum vectors and the large-scale structures from the Sloan survey and claimed that their results 
are in qualitative agreement with the predictions of the tidal torque theory. \citet{slo-etal09} reported 
a first detection of the chiral correlations of nearby galaxies.
Recently, \citet{jones-etal10} found that the spin axes of the spiral galaxies in the filaments 
tend to be aligned with the major axes of the filaments and concluded that the observed spin 
alignments of the filament galaxies provide a fossil evidence for the tidally-induced intrinsic 
correlations. 

In spite of these observational evidences, it is still inconclusive whether or not 
the present galaxies still retain its initially memory of the tidally induced 
intrinsic spin correlations for the following two reasons. The first reason is that the 
reported correlation signals are not strong enough to be confirmed as true signals. 
For instance, the galaxy spin correlation signals reported by \citet{pen-etal00}, 
\citet{LP01} and \citet{LE07} are significant only at $2-2.5\sigma$ levels. The samples 
on which the analyses of \citet{nav-etal04} and \citet{tru-etal06} were based are so
small that their conclusions suffer from poor number statistics. Furthermore, there have 
reported some counter evidences against the existence of the intrinsic galaxy alignments. For 
example, \citet{AS05} searched for the alignments of the spin vectors of the galaxies in Abell 
clusters but failed in finding any signals.

As for the strong correlations of the projected major axes of the elliptical galaxies measured by 
\citet{brown-etal02}, although it is good number statistics, the signals suffer from large 
uncertainties associated with regarding the projected major axes of the elliptical galaxies as the 
directions orthogonal to the spin axes of the underlying dark halos. True as it is that the 
minor axes of the elliptical galaxies should be more or less aligned with the spin axes of 
their underlying dark halos \citep[e.g.,][]{bai-etal05,BS05}, their spatial correlations 
are not so good tracers of the linear tidal fields as the spin correlations of the spiral 
galaxies, since the shapes of the elliptical galaxies are apt to be affected by the subsequent 
processes such as the anisotropic merging/infall, galaxy-galaxy interactions and etc 
\citep[e.g.,][]{ful-etal99}. Thus, the best targets for the measurement of the tidally-induced 
intrinsic alignments are the spiral galaxies and their spin orientations.

The second reason comes from the concerns about spurious signals that could be produced by 
systematics. There are three major sources for the systematics in the measurement of the 
intrinsic  correlations of the galaxy spin axes: the weak gravitational lensing effect, 
presence of thick central bulges and inaccurate measurements of the spin axes in 
case that the galaxies have small angular sizes. It has been known that the extrinsic alignments 
of the galaxy shapes due to the weak gravitational lensing effect would create spurious signals of 
the spin-spin correlations of the spiral galaxies (or ellipticity-ellipticity correlations of the 
elliptical galaxies) at redshifts $z\ge 0.1$ \citep[e.g.,][]{pen-etal00,CM00,hea-etal00,cat-etal01,cri-etal01,
cri-etal02,jin02,mac-etal02,HH03,HS04,hir-etal04,man-etal06,hir-etal07,JB10}. 
The presence of thick central bulges in the spiral galaxies could cause significant systematics since it invalidates 
the circular thin-disk approximation which is almost exclusively used to measure the orientations of the galaxy 
spin axes. In case that the galaxies have small angular sizes, their position angles and axial ratios would be 
difficult to measure accurately, which would propagate into the systematic errors in the measurement of the 
galaxy spin axes.

The errors caused by the above three systematics in the measurement of the intrinsic correlations 
can be minimized if one selects only the large late-type spiral galaxies (of Hubble types Scd) observed 
at lowest redshifts. Using only low-$z$ galaxies, one can reduce the weak gravitational lensing effect to a 
completely negligible level. Selecting only large late-type spiral galaxies which have the smallest central bulges, 
one can guarantee that the measurements of the spin axes through the circular thin disk approximation are reliable. 
To select such galaxies from the observational data, however, it is necessary to have information on the 
spectroscopic redshifts and Hubble types of the observed galaxies.  
Very recently, \citet{HC-etal10} have released a catalog of the galaxies from the Sloan Digital Sky Survey Data 
Release 7 (SDSS DR7), in which such information are all available. 
Our goal here is to measure the true intrinsic correlations of the nearby large late-type spiral galaxies 
from this catalog and to study their behaviors and their dependence on the local density and luminosity. 

The organization of this Paper is as follows. In \S 2.1, the sample of the nearby large late-type galaxies selected 
from the SDSS DR7 is described and the number distributions of the selected galaxies as a function of their 
magnitude and local density are derived. In \S 2.2, it is explained how the spin axes of the selected galaxies 
are determined up to two-fold degeneracy with the help of the circular thin disk approximation. 
In \S 3.1, the correlations between the spin axes of the selected nearby large Scd galaxies 
are measured and the bootstrap error analysis is presented. In \S 3.2 the dependence of the galaxy spin 
correlations on local density and luminosity is investigated. In \S 3.3 the possible residual systematics are 
discussed and its effect is examined. In \S 4, the observational results are compared with the analytic models 
based on the tidal torque theory and the bestfit parameters to quantify the strengths of the correlations are 
determined.In \S 5, the results are summarized and a conclusion is drawn.

\section{PROPERTIES OF THE LATE-TYPE SPIRAL GALAXIES FROM SDSS DR7}

\subsection{Selection of the Nearby Large Scd Galaxies}

The galaxy catalog compiled by \citet{HC-etal10} consists of a total of $698420$ galaxies 
in redshift range of $0\le z\le 0.16$ from the SDSS DR7 \citep{sdssdr7}. 
It provides information on spectroscopic redshift ($z$), right ascension ($\alpha$), 
declination ($\delta$) and the probabilities of being in the five  morphological classes 
(E,\ Ell,\ S0,\ Sab,\ Scd) for each galaxy. Through private communication with M. Huestra-Company, 
we obtain additional information on the major and minor axes, position angle ($\vartheta_{P})$ and r-band 
model magnitude $(m_{r}$) of each galaxy in the catalog.  

\begin{figure}
\begin{center}
\plotone{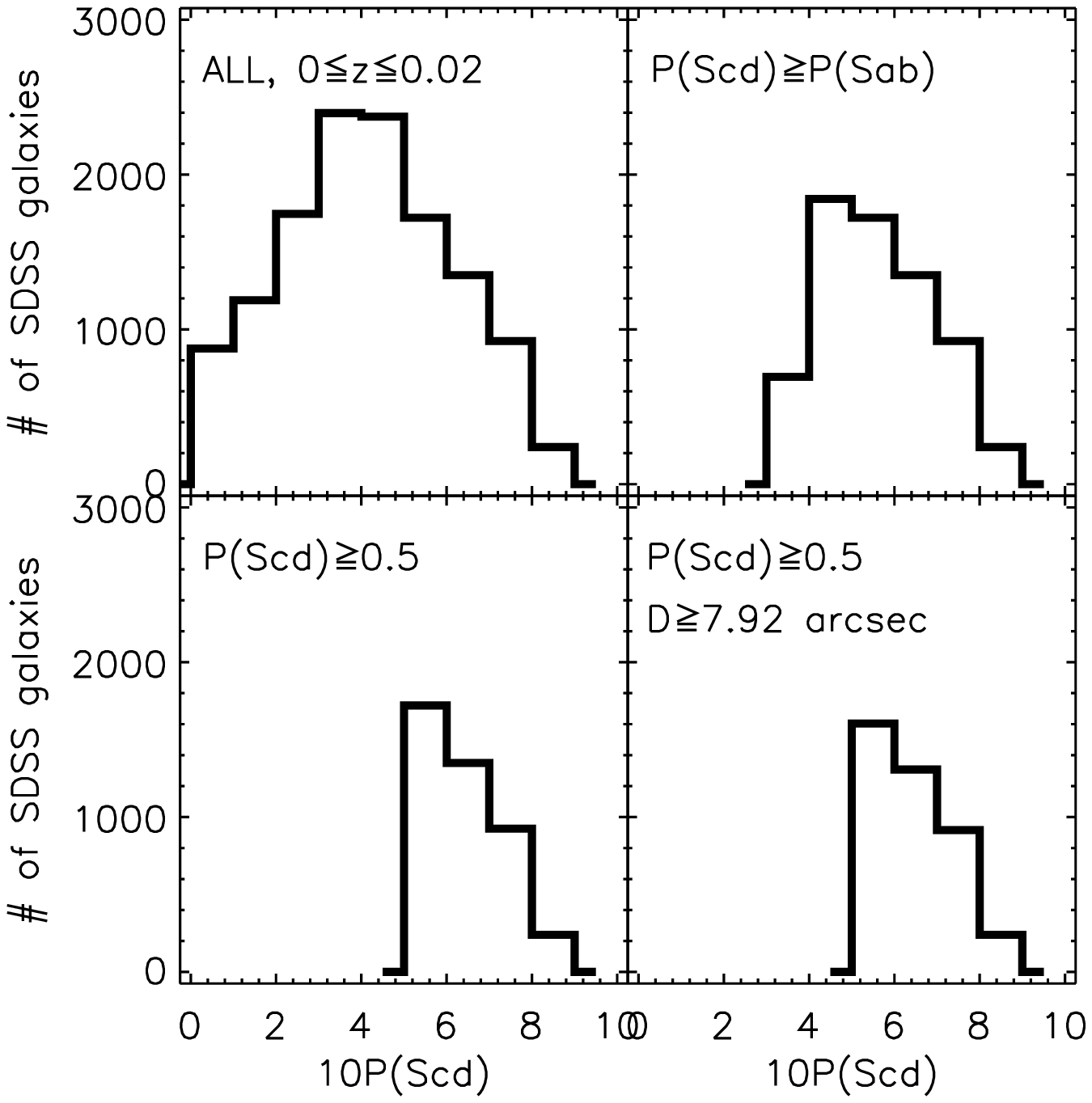}
\caption{Number distribution of the nearby galaxies ($0\le z\le 0.02$) from the SDSS 
DR7 as a function of $P(Scd)$, the probability of being classified as Scd galaxies for 
four different cases: all nearby galaxies (top left panel); $P(Scd)$ is largest (top right panel);  
$P(Scd)\ge 0.5$ (bottom left panel);  with diameter larger than $20h^{-1}$Kpc and 
$P(Scd)\ge 0.5$ (bottom right panel).}
\label{fig:dis_pscd}
\end{center}
\end{figure}

To minimize the systematics in the measurement of the spatial correlations of the galaxy spin axes, 
we select only those galaxies which have redshifts in the range of $0\le z\le 0.02$, probability 
of being in the morphological class of Scd higher than $0.5$, and angular size (given as the major 
axis length) $D$ larger than $7.92$ arcsec (corresponding to 20 pixels in the SDSS frames).
A total of $4065$ galaxies from the catalog are found to satisfy these three conditions. 
Figure \ref{fig:dis_pscd} plots the number distribution of the galaxies from the catalog as 
a function of its probability of being in the Scd morphological class, $P({\rm Scd})$. The results 
shown in each panel is drawn under the condition of $0\le z\le 0.02$ (top left panel); $0\le z\le 0.02$ 
and $P({\rm E})\le P({\rm S0})\le P({\rm Sab})\le P({\rm Scd})$ (top right panel); $0\le z\le 0.02$ and 
$P({\rm Scd})\ge 0.5$ (bottom left panel); $0\le z\le 0.02$, $P({\rm Scd})\ge 0.5$ and $D\ge 7.92$ arcsec 
(bottom right panel).  Using information on $z$, $\alpha$ and $\delta$, we determine the comoving distance to 
each galaxy in unit of $h^{-1}$Mpc assuming a WMAP7 cosmology \citep{wmap7}.  

\begin{figure}
\begin{center}
\plotone{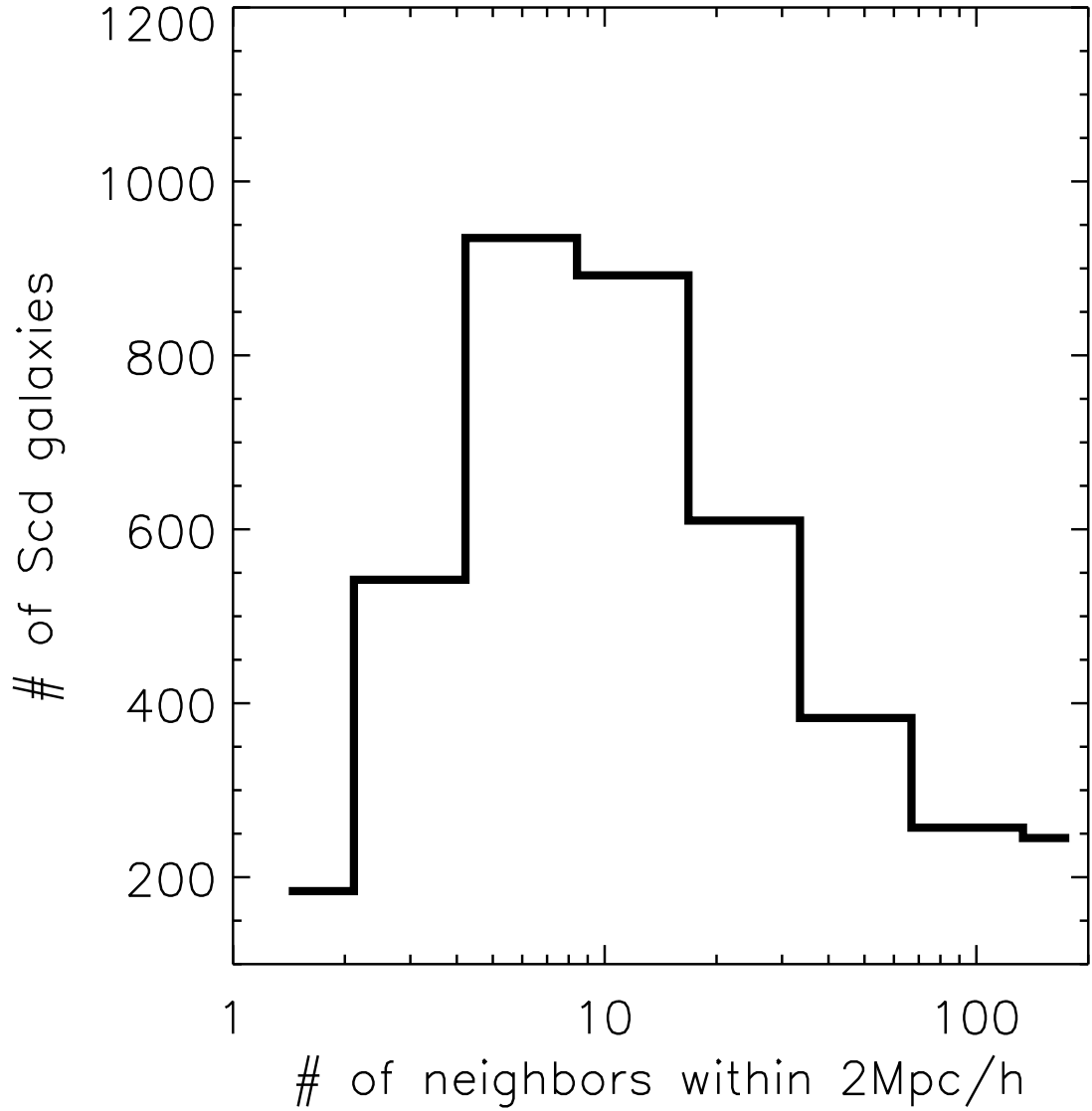}
\caption{Number distribution of the selected Scd galaxies from the SDSS DR7 as a function of the 
number of neighbor galaxies located within the separation distance of $2\,h^{-1}$Mpc at 
$0\le z < 0.02$.}
\label{fig:dis_den}
\end{center}
\end{figure}

\begin{figure}
\begin{center}
\plotone{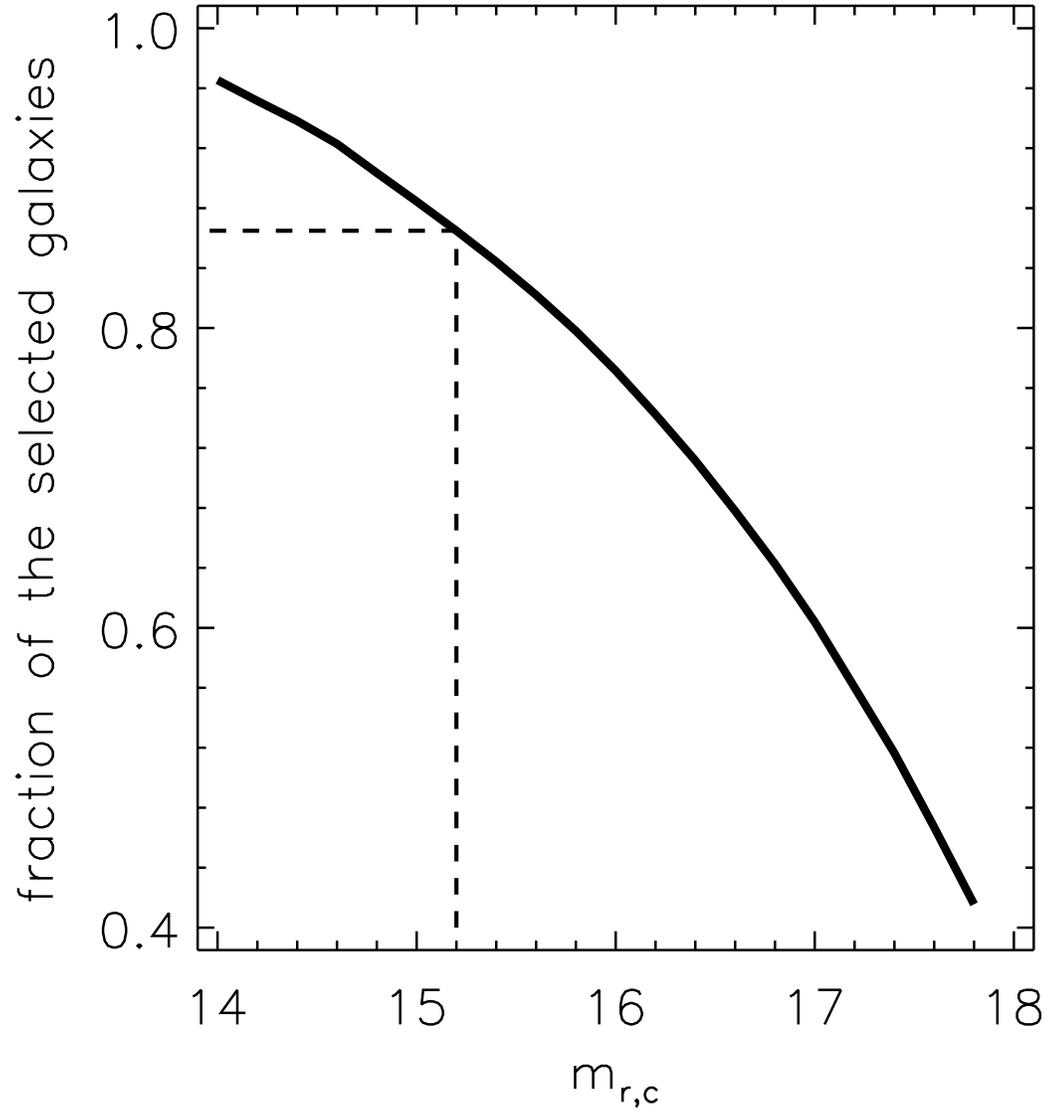}
\caption{Fraction of the galaxies belonging to the volume-limited SDSS sample as a function of the $r$-band 
magnitude limit, $m_{r,c}$. The dashed line corresponds to the applied magnitude cut.}
\label{fig:vcut}
\end{center}
\end{figure}

To determine the local density of the selected galaxies,  we first construct a volume-limited sample of the 
galaxies (regardless of their types) in the same redshift range from the SDSS spectroscopic data. 
Basically, the volume-limited sample includes only those SDSS galaxies whose apparent $r$-band 
magnitudes would exceed a given flux limit, $m_{r,c}$, if placed at $z=0.02$. Figure \ref{fig:vcut} plots the fraction 
of the SDSS galaxies included in a volume-limited sample as a function of $m_{r,c}$.  When the SDSS flux limit value 
of $m_{r,c}=15.2$ is applied, a total of $12273$ galaxies are found to belong to the constructed volume-limited 
sample. Now, for each selected nearby Scd galaxy from the catalog provided by \citet{HC-etal10}, we count the 
number of its neighbor galaxies from the volume-limited sample whose separation distance from 
the given selected Scd galaxy is less than $r_{s}=2\,h^{-1}$Mpc. Figure \ref{fig:dis_den} plots the 
number distribution of the selected nearby large Scd galaxies as a function of the neighbor galaxies located 
within separation of $r_{s}$. The distribution has its maximum at $N_{ng}\approx 10$ and only a small 
fraction of the selected Scd galaxies have indeed more than $100$ neighbor galaxies within $r_{s}$. 
It indicates that the majority of the selected nearby Scd galaxies are field galaxies, not belonging to galaxy clusters.

\begin{figure}
\begin{center}
\plotone{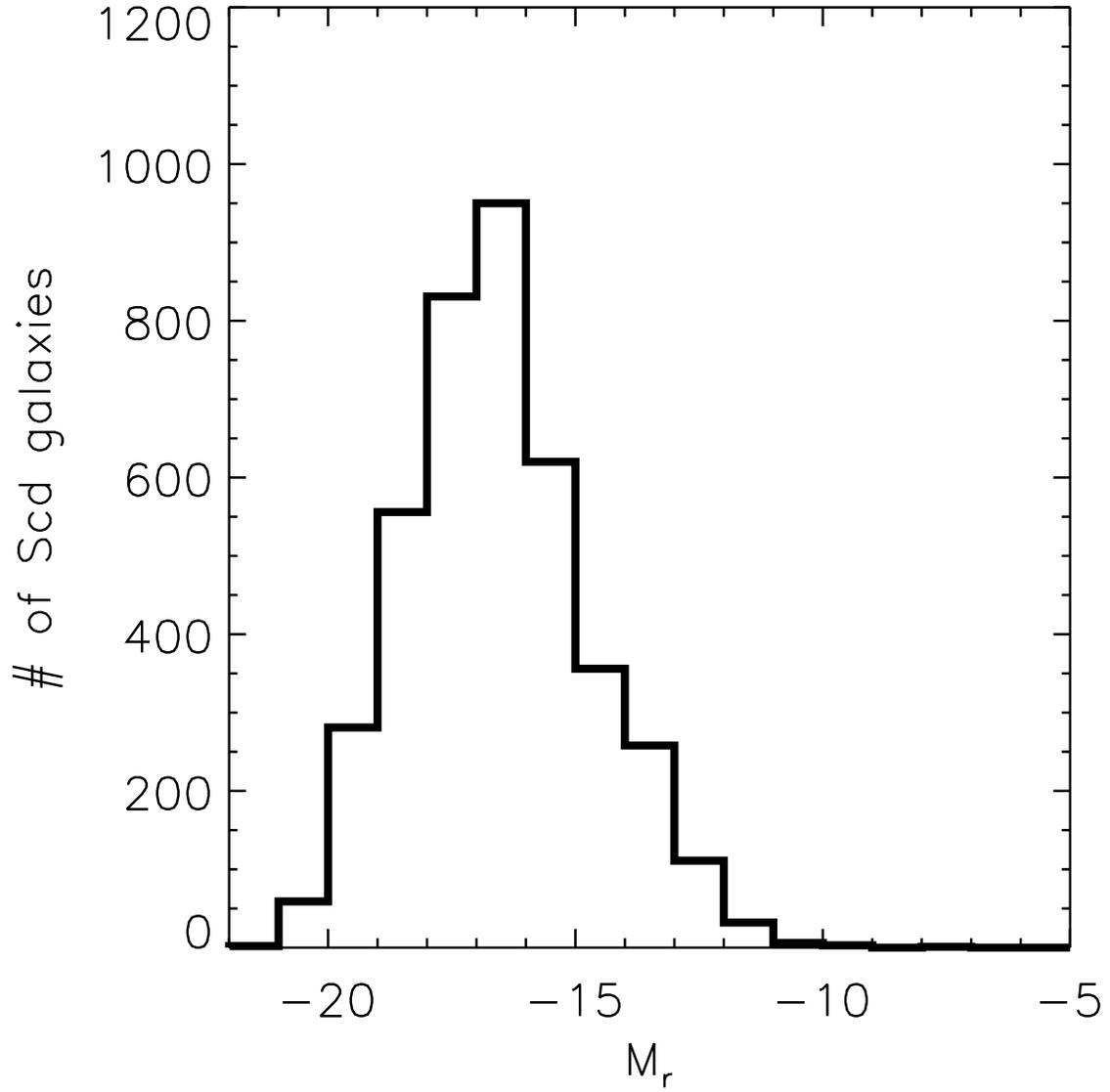}
\caption{Number distribution of the selected Scd galaxies from the SDSS DR7 as a function of the 
r-band absolute magnitude at $0\le z < 0.02$. }
\label{fig:dis_mag}
\end{center}
\end{figure}

With the measured comoving distance to each galaxy, we convert each galaxy's apparent magnitude 
$m_{r}$ to the absolute magnitude $M_{r}$ by using the system of the inverse hyperbolic sine magnitudes 
(asinh)  \citep{lup-etal99}. Figure \ref{fig:dis_mag} plots the number distribution of the selected nearby 
large Scd galaxies as a function of $M_{r}$.  Table \ref{tab:Scd} lists the redshift range, the total 
number, mean absolute r-band magnitude and mean number of the neighbors within $2\,h^{-1}$Mpc, 
averaged over all the selected nearby large Scd galaxies.

\subsection{Determination of the Galaxy Spin Axes}

To determine the spin axes of each selected galaxy, we use information from the SDSS imaging data on galaxy's axial 
ratio $q$ and position angle $\vartheta_{P}$ which were extracted by the SDSS team from the coefficients of the 
Fourier expansion of the 25 magnitudes per square arcsecond isophote measured in the $r$- band.  
Placeholders for the errors on each of these quantities were not available from the website (http://www.sdss.org/dr7). 
We calculate the inclination angle, $\xi$, of each selected galaxy as \citep{HG84}
\begin{equation}
\label{eqn:xi}
\cos^{2}\xi = \frac{q^{2}-p^{2}}{1-p^{2}}.
\end{equation}
where $p$ is the intrinsic flatness parameter that depends on the galaxy morphological 
type. Here, we adopt the value of $p=0.10$ for the Scd galaxies, given by \citet{HG84}.  
If the inclination angle of a given galaxy in the selected sample is less than the intrinsic flatness 
parameter (i.e., $q \le p$), then we set it at $\pi/2$ (see \S 3.3 for discussion on the systematics 
related to the intrinsic flatness parameter).

With the help of the circular thin-disk approximation and given information on the 
inclination angle $\xi$ and position angle $\vartheta_{P}$, we determine the unit spin 
vector, $\hat{\bf L}$, of each selected nearby Scd galaxy in a local spherical-polar 
coordinate system up to sign-ambiguity of the radial component as \citep{LE07}
\begin{eqnarray}
\hat{L}_{r}&=&\pm\cos\xi, \\ 
\hat{L}_{\theta}&=&(1-\cos^{2}\xi)^{1/2}\sin\vartheta_{P},\\ 
\hat{L}_{\phi}&=&(1-\cos^{2}\xi)^{1/2}\cos\vartheta_{P} 
\end{eqnarray}
where $\hat{L}_{r},\ \hat{L}_{\theta},\ \hat{L}_{\phi}$ correspond to the radial, polar 
and azimuthal component of $\hat{\bf L}$, respectively. This sign-ambiguity in the 
radial components of $\hat{\bf L}$ is due to the fact that one cannot determine whether 
the rotation of a given galaxy upon its symmetry axis is clock-wise or counter-clock 
wise \citep{pen-etal00}. 

Using information on the equatorial coordinates, $(\alpha,\ \delta)$, we determine the 
unit spin vector of each selected Scd galaxy in the Cartesian coordinate system up to 
two-fold degeneracy as 
\begin{eqnarray}
\hat{L}_{a1}&=&\hat{L}_{r}\sin\theta\cos\phi + \hat{L}_{\theta}\cos\theta\cos\phi  - 
\hat{L}_{\phi}\sin\phi ,\\
\hat{L}_{a2} &=&\hat{L}_{r}\sin\theta\sin\phi  + \hat{L}_{\theta}\cos\theta\sin\phi + 
\hat{L}_{\phi}\cos\phi ,\\
\hat{L}_{a3} &=& \hat{L}_{r}\cos\theta - \hat{L}_{\theta}\sin\theta , 
\end{eqnarray}
\begin{eqnarray}
\hat{L}_{b1}&=&-\hat{L}_{r}\sin\theta\cos\phi + \hat{L}_{\theta}\cos\theta\cos\phi  - 
\hat{L}_{\phi}\sin\phi ,\\
\hat{L}_{b2} &=&-\hat{L}_{r}\sin\theta\sin\phi  + \hat{L}_{\theta}\cos\theta\sin\phi + 
\hat{L}_{\phi}\cos\phi ,\\
\hat{L}_{b3} &=& -\hat{L}_{r}\cos\theta - \hat{L}_{\theta}\sin\theta , 
\end{eqnarray}
where $\theta=\pi/2-\delta$ and $\phi=\alpha$. Finally, to each selected nearby Scd galaxy, 
we assign a set of two unit spin vectors, $\hat{\bf L}_{a}$ and $\hat{\bf L}_{b}$,  
which differ from each other by the sign of $\hat{L}_{r}$ \citep{pen-etal00}. 

\section{INTRINSIC SPIN CORRELATIONS OF THE LATE-TYPE SPIRAL GALAXIES}

\subsection{Measurement of the Galaxy Spin Correlations}

The spatial correlation of the galaxy spin axes is defined by \citet{pen-etal00} as
\begin{equation}
\label{eqn:eta}
\eta(r) \equiv \langle\vert\hat{\bf L}({\bf x})\cdot
\hat{\bf L}({\bf x + r})\vert^{2}\rangle - \frac{1}{3}.
\end{equation}
Here the ensemble average is taken over those galaxy pairs whose separation distance is 
in a differential range of $[r,\ r +d r]$. The value of $1/3$ is subtracted since it is 
the value of the ensemble average when there is no correlation.  
In practice, however, due to the two-fold degeneracy in the determination of $\hat{\bf L}$,  
the correlation $\eta(r)$ can be measured only as  \citep{pen-etal00} 
\begin{equation}
\label{eqn:prac}
\eta(r) = \frac{1}{4}\left(\langle \vert\hat{\bf L}_{a}\cdot\hat{\bf L}^{\prime}_{a}\vert^{2}\rangle + 
\langle \vert\hat{\bf L}_{a}\cdot\hat{\bf L}^{\prime}_{b}\vert^{2}\rangle + 
\langle \vert\hat{\bf L}_{b}\cdot\hat{\bf L}^{\prime}_{a}\vert^{2}\rangle + 
\langle \vert\hat{\bf L}_{b}\cdot\hat{\bf L}^{\prime}_{b}\vert^{2}\rangle\right) - \frac{1}{3}
\end{equation}
where $\hat{\bf L}^{\prime}$ represents the unit spin vector measured at 
${\bf x}+{\bf r}$. 

For all pairs of the galaxies, we calculate the squares of the dot products of their unit spin vectors, 
taking into account the two-fold degeneracy. Binning the separation distances $r$, we take the 
ensemble average over those pairs whose separation distances  belong to a certain distance bin and 
subtract $1/3$ to determine $\eta(r)$ in accordance with Equation (\ref{eqn:prac}). 
Although we try to reduce systematic errors as much as possible through constraining the redshift-range, 
morphology and angular size, there could be some residual systematics. To sort out possible 
residual systematics, we perform a bootstrap error analysis.  The spin correlation functions, $\eta(r)$, 
are remeasured from each of the $1000$ bootstrap resamples which is constructed through  shuffling 
randomly the positions of the selected galaxies.  The bootstrap errors, $\sigma_{b}$, are calculated as 
$\sigma_{b}\equiv\langle(\eta-\bar{\eta}_{b})^{2}\rangle^{1/2}$ where the ensemble 
average is taken over the $1000$ resamples and $\bar{\eta}_{b}$ represents the bootstrap mean. 
If there were no residual systematics in the measurement of the galaxy spin axes, then the bootstrap mean value 
would be very close to zero. The degree of the deviation of the bootstrap mean from zero would indicate the level 
of the residual systematics.

\begin{figure}
\begin{center}
\plotone{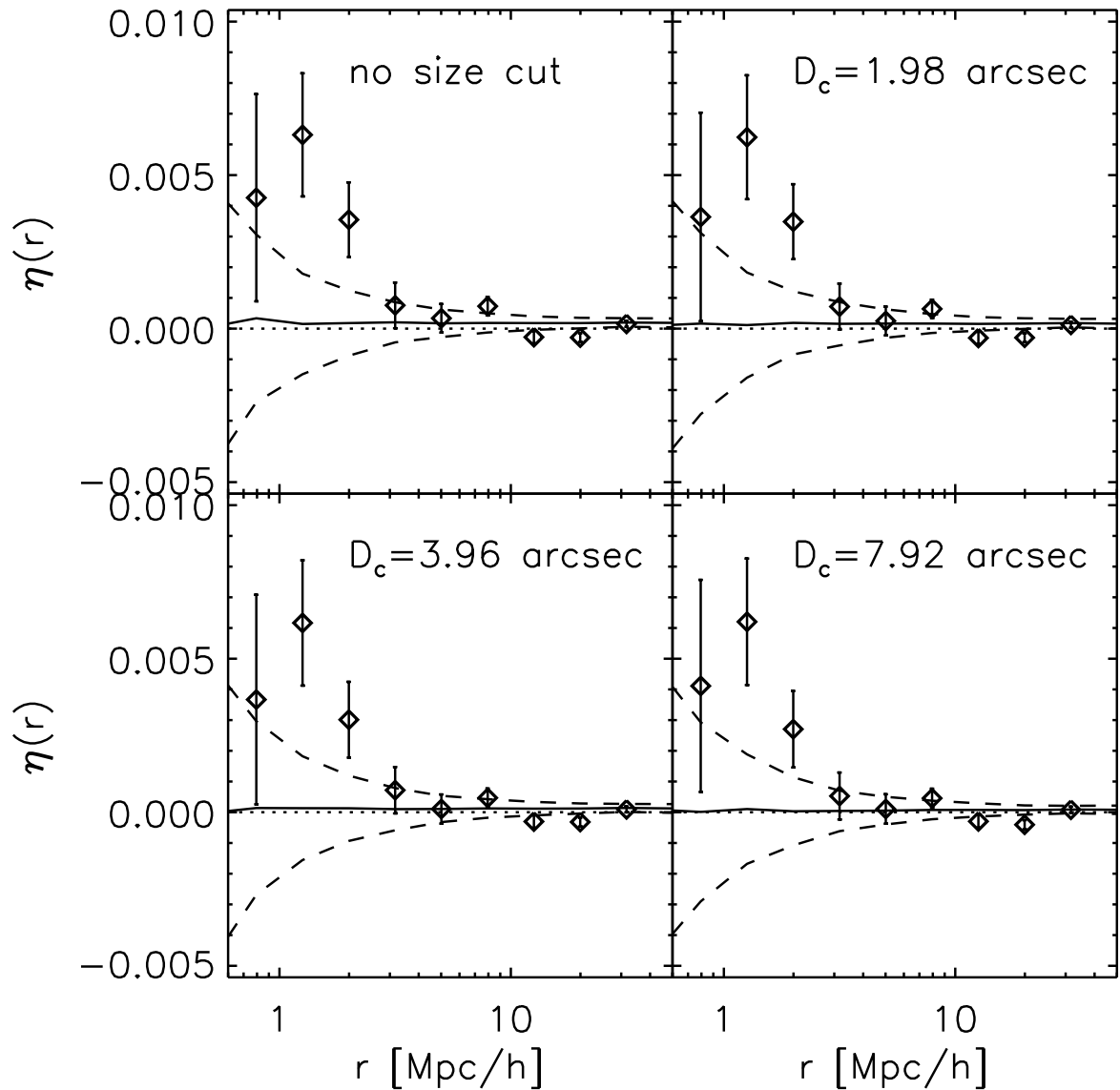}
\caption{Correlations of the unit spin vectors  of the selected Scd galaxies (diamond dots) as a function of 
the separation distance $r$ at $0\le z < 0.02$ for the four different cases of the galaxy diameter cut, $D_{c}$. 
In each panel, the thin solid line represents the mean correlation averaged over the $1000$ bootstrap resamples, 
the dashed lines represent one standard deviation  among the bootstrap resamples, and the dotted line 
indicates the zero correlation signal.}
\label{fig:spincor_z}
\end{center}
\end{figure}

Figure \ref{fig:spincor_z} plots the observed galaxy spin correlations for four different cases of the galaxy 
angular size cut  ($D_{c}=0.00,\ 1.98,\ 3.96$ and $7.92$ arcsec in the top-left, 
top-right, bottom-left and bottom-right panel, respectively). In each panel the error bars represent one standard 
deviation, $\sigma_{\eta}$, in the measurement of $\eta(r)$.  
The thin solid and dashed line in Figure \ref{fig:spincor_z} represents the bootstrap mean 
$\bar{\eta}_{b}$ and bootstrap errors, $\sigma_{b}$, respectively, while the thin dotted line 
corresponds to the zero signal.  As it can be seen, a clear signal as strong as $3.4\sigma_{b}$ is 
detected for all four cases of $D_{c}$ at the second radial bin corresponding to the distance of 
$r\approx 1.25 h^{-1}{\rm Mpc}$. For the cases of $D_{c}=0$ and $1.98$ arcsec, the third radial bins 
corresponding to $r\approx 2\,h^{-1}$Mpc also exhibit a $3\sigma_{b}$ signal while for the other two 
cases of $D_{c} =3.96$ and $7.92$, the correlation at $r=2\,h^{-1}$Mpc diminishes down to $2.4\sigma_{b}$ 
level. 

\begin{figure}
\begin{center}
\plotone{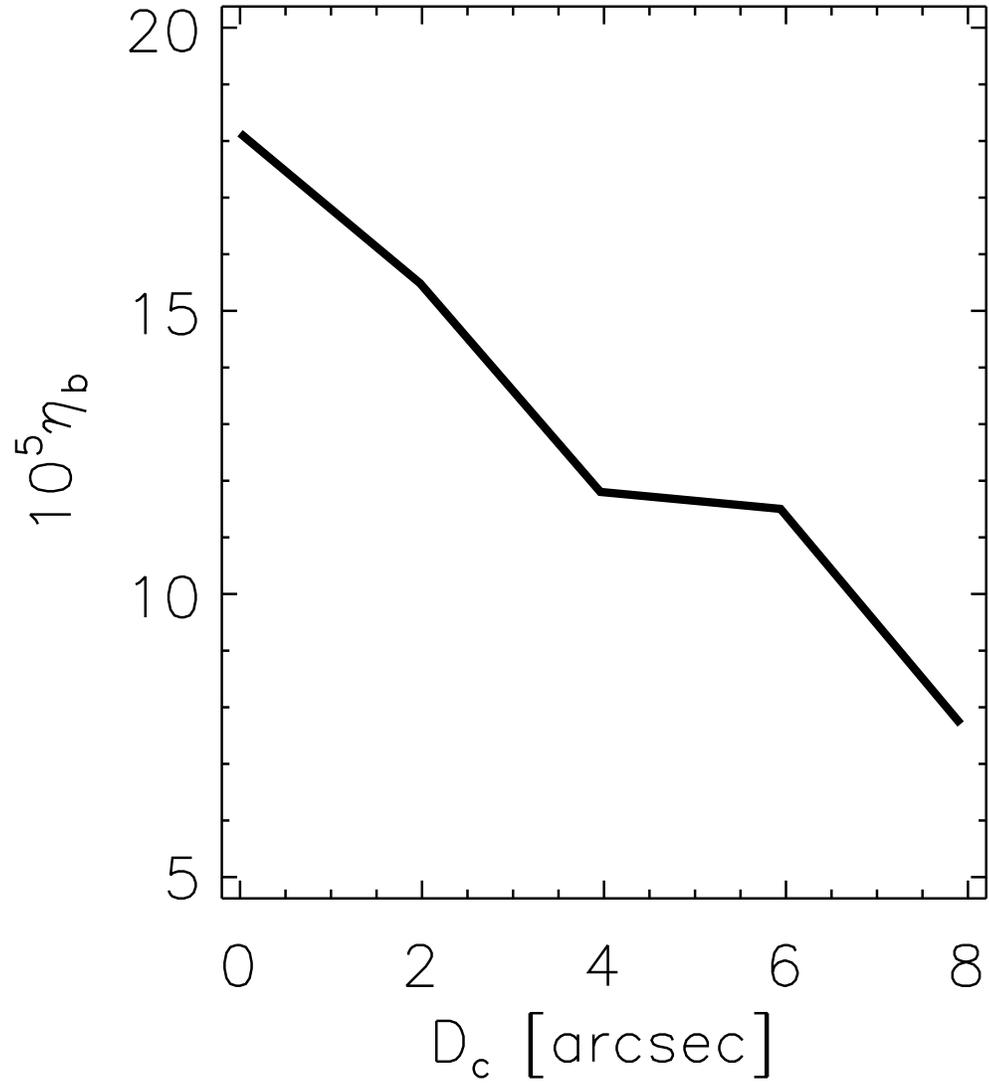}
\caption{Bootstrap mean averaged over the $1000$ bootstrap resamples measured at 
$r\sim 10\,h^{-1}$Mpc, as a function of $D_{c}$.}
\label{fig:booterr}
\end{center}
\end{figure}

As mentioned in \S 1, for the galaxies having small angular sizes the spin axes are difficult to measure 
accurately. As it can be seen from Figure \ref{fig:spincor_z}, for the two cases of  $D_{c}=0$ and $1.98$ arcsec 
where the small angular size galaxies are included,  the mean bootstrap values (thin solid line) show non-negligible 
deviations from zero (larger than $10^{-4}$). Thus, the $3\sigma_{b}$ signals detected 
at $r\approx 2\,h^{-1}$Mpc for these two cases are likely to be contaminated by the systematics. 
Figure \ref{fig:booterr} shows more clearly how the averaged bootstrap mean value changes with 
$D_{c}$. As it can be seen, the bootstrap mean value decreases  as $D_{c}$ increases and drops below 
$10^{-4}$ (negligible level) when $D_{c}$ reaches up to $7.92$ arcsec, which justify our choice of 
$D_{c}=7.92$ arcsec for this analysis. 

Since the correlation function, $\eta(r)$, is computed as a function of the three dimensional separation, 
there may be some cross-correlations at different radii.  If the cross-correlations between the second and 
third radial bins are significant, then the signal detection for the second radial bin would have lower 
significance.  To quantify the amount of cross-correlations, we compute the covariance matrix, $C_{ij}$, that is 
defined 
\begin{equation}
\label{eqn:cov}
C_{ij}\equiv\langle (\eta_{i}-\bar{\eta}_{bi})(\eta_{j}-\bar{\eta}_{bj})\rangle
\end{equation}
where the ensemble average is taken over $1000$ bootstrap resamples, $\eta_{i}$ is the correlation at the 
$i$-th bin from a bootstrap resample, and $\bar{\eta}_{bi}$ is the bootstrap mean at the $i$-th bin.  
If the cross-correlations between the second and third radial bins are significant, then the off-diagonal 
component, $C_{23}$, would be comparable in magnitude to the diagonal component, $C_{22}$. 
It is found that $C_{22}=3.18\times 10^{-6}$ while $C_{23}=-0.17\times 10^{-7}$. This implies that  
the cross-correlations between the second and third radial bins are order of magnitude smaller (and even negative) 
than the correlations at the second radial bins, which proves that the signal detected at the second radial 
bin is robust.

\subsection{Dependence on Local Number Density and Luminosity}

\begin{figure}
\begin{center}
\plotone{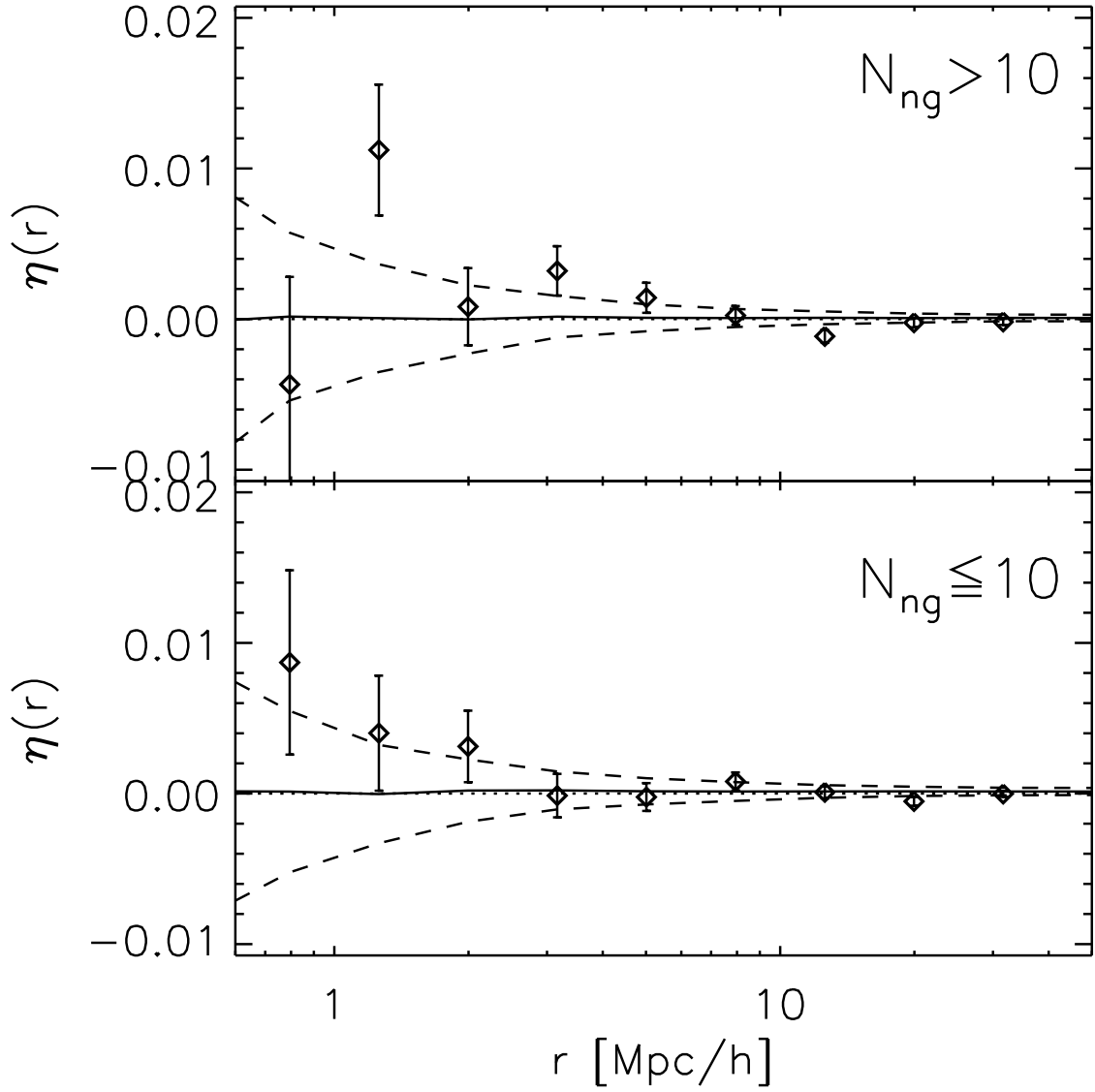}
\caption{Same as Figure \ref{fig:spincor_z} but using only those Scd galaxies which 
have more (less) than $N_{ng,c}=10$ neighbor galaxies within separation $r_{s}=2\,h$Mpc 
in the upper (lower) panel.}
\label{fig:spincor_den}
\end{center}
\end{figure}

Now that we have detected non-zero galaxy spin correlations from the SDSS data, 
we would like to investigate whether or not $\eta(r)$ depends on the local number 
density $N_{ng}$ of neighbor galaixses. Applying the number density cut, $N_{ng,c}=10$, to 
the sample, we divide the selected Scd galaxies into two subsamples each of which consists of 
those nearby large Scd galaxies with $N_{ng}\le N_{ng,c}$ and $N_{ng}>N_{ng,c}$, respectively. 
Then, we measure $\eta(r)$ from each subsample separately, the results of which are plotted
in the top and bottom panel of Figure \ref{fig:spincor_den}, respectively. 

As it can be seen, for the subsample with $N_{ng}>10$, we find a correlation signal as significant as 
$3.1\sigma_{b}$ signal at $r\approx 1.25\,h^{-1}$Mpc, while for the subsample with $N_{ng}\le 10$, 
no correlation signal is detected. The bootstrap mean values are quite close to zero for both of the cases, 
which indicates that the detected signal is not spurious. This result shows that those Scd galaxies located 
in denser regions tend to have stronger spin correlations.  A possible physical explanation is that the galaxies 
located in denser regions experience so stronger tidal forces from the surrounding neighbor mass distribution 
that the tidally-induced intrinsic correlations of their spin axes are better retained, which is consistent with the 
previous results of \citet{LE07}. 

\begin{figure}
\begin{center}
\plotone{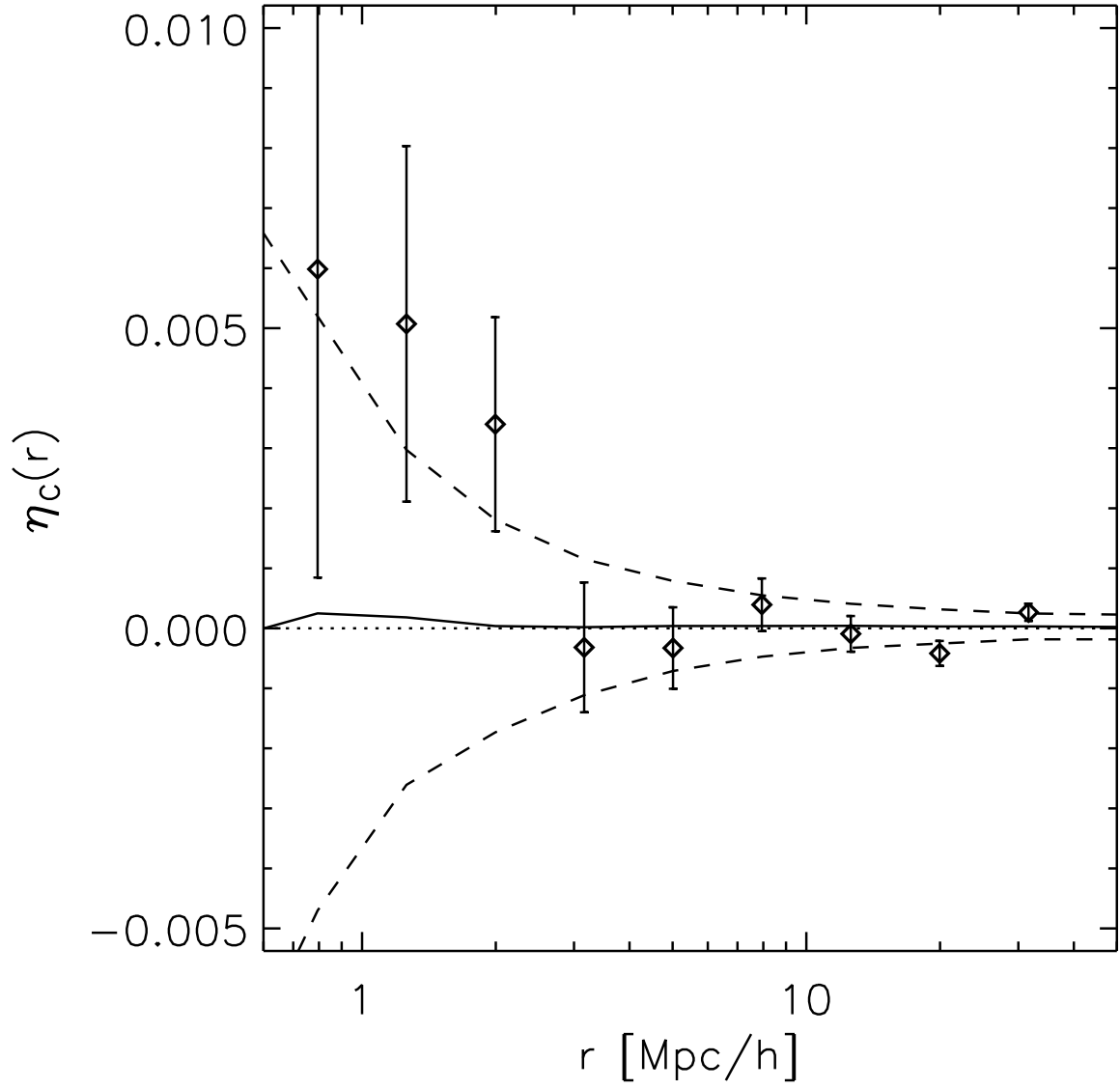}
\caption{Same as Figure \ref{fig:spincor_crossden} but between the Scd galaxies having 
more than $10$ neighbors and the Scd galaxies having less than $10$ neighbors.}
\label{fig:spincor_crossden}
\end{center}
\end{figure}

Using the galaxy pairs in which one galaxy has $N_{ng}>N_{ng,c}$ and the other has 
$N_{ng}\le N_{ng,c}$, we also measure the cross-correlations of the spin axes between 
the two subsamples, the result of which is plotted in Figure \ref{fig:spincor_crossden}. 
As one can see, the result is consistent with zero cross-correlation, which reveals that 
the spin axes of the galaxies in less dense regions are not correlated with those 
of the galaxies in denser regions. 

\begin{figure}
\begin{center}
\plotone{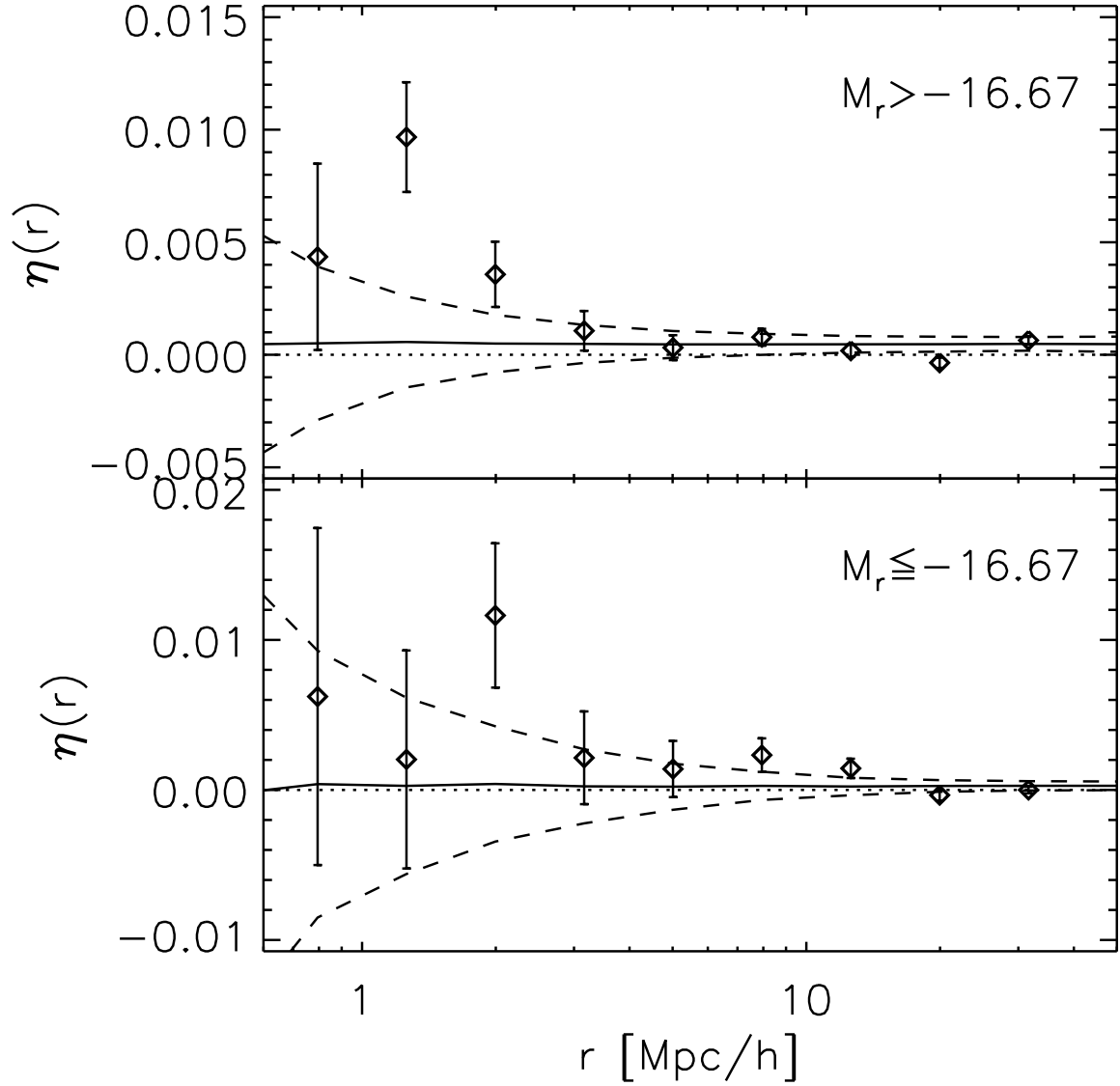}
\caption{Same as Figure \ref{fig:spincor_z} but using only those Scd galaxies which are 
fainter (brighter) than $M_{r,c}=-16.67$ in the upper (lower) panel. }
\label{fig:spincor_mag}
\end{center}
\end{figure}

To investigate  how the intrinsic correlations depend on the luminosity, we take 
the median $r$-band absolute magnitude, $-16.67$, as the threshold $M_{r,c}$, and 
divide the selected galaxies into two subsamples each of which consists of those nearby 
large Scd galaxies with $M_{r}>-16.67$ and $M_{r}\le -16.67$, respectively. Then, we measure 
$\eta(r)$ from each subsample separately, which are plotted in the top and bottom panel of 
Figure \ref{fig:spincor_mag}, respectively. As it can be seen, the correlation function, $\eta(r)$, 
exhibits  a $4.9\sigma_{b}$ peak at $r\approx 1.25\,h^{-1}$Mpc  and a $2.4\sigma_{b}$ peak at 
$r\approx 2\,h^{-1}$Mpc in the fainter and brighter galaxy sample, respectively. 

\begin{figure}
\begin{center}
\plotone{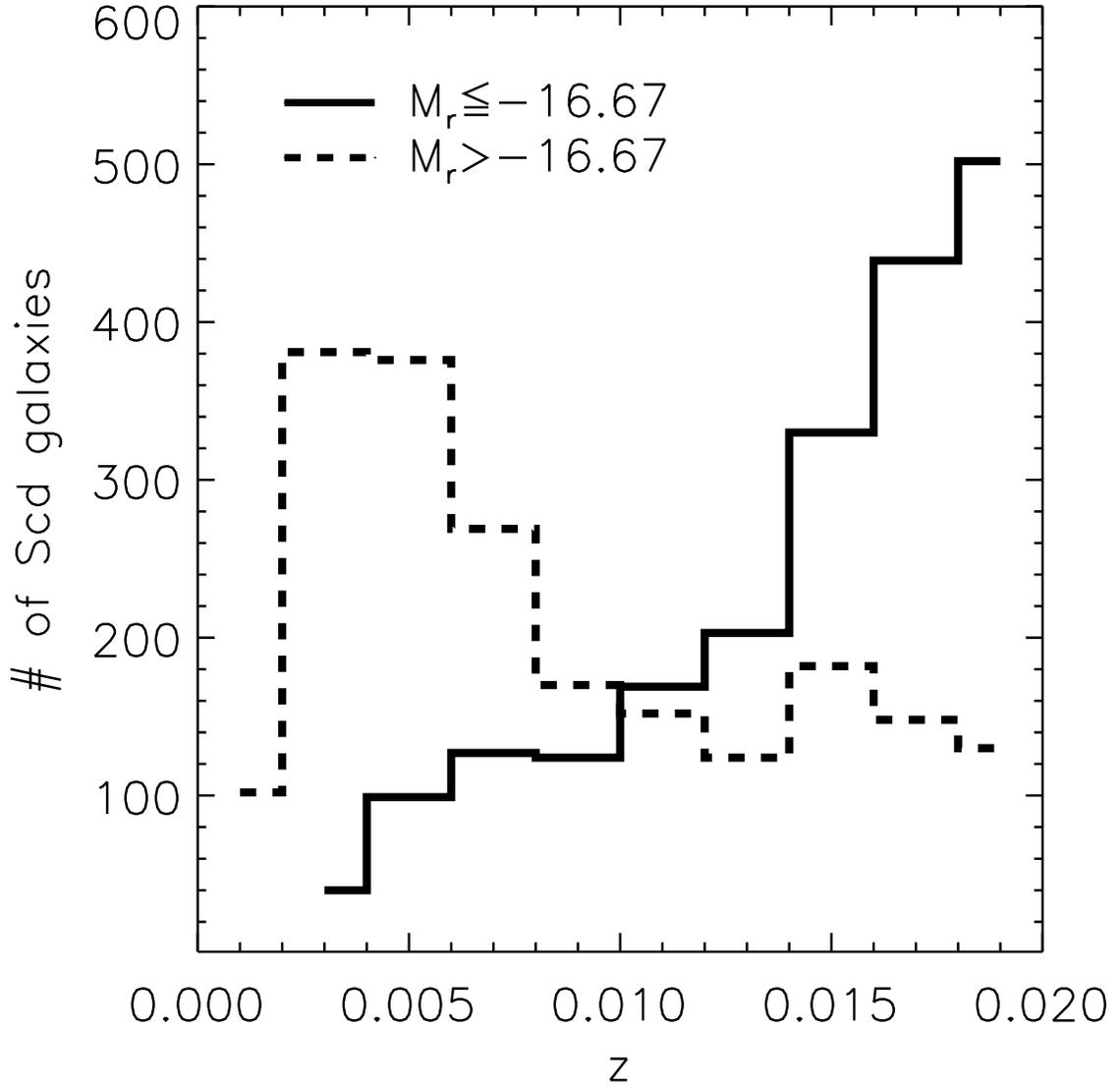}
\caption{Redshift distribution of the fainter galaxy sample (dashed line) and bright galaxy sample (solid line).}
\label{fig:zdis}
\end{center}
\end{figure}

This result, however, cannot be interpreted as a clear luminosity dependence of the galaxy spin 
correlations. In fact, the peaks of $\eta(r)$ detected at $1.25\,h^{-1}$Mpc and $2\,h^{-1}$Mpc 
(for the fainter and brighter sample, respectively) turn out to correspond to a similar angular 
separation of galaxy pairs. It indicates the existence of significant residual systematics, which is also manifest 
from the relatively high degree of the deviation of the bootstrap mean from zero for the fainter sample as 
shown in the top panel of Figure \ref{fig:spincor_mag}. Furthermore, the redshift distributions of the two 
subsamples are found to be largely different: the brighter (fainter) sample is biased to relatively higher 
(lower) redshifts, as shown in Figure \ref{fig:zdis}. The mean and median redshifts of the two subsamples 
are also listed in Table \ref{tab:zdis}.  Given that there should exist significant systematics and that it is impossible 
to disentangle a luminosity from a redshift dependence of the galaxy spin correlations, we admit that a luminosity 
dependence of the galaxy spin correlations cannot be measured with our magnitude-split samples. 

\begin{figure}
\begin{center}
\plotone{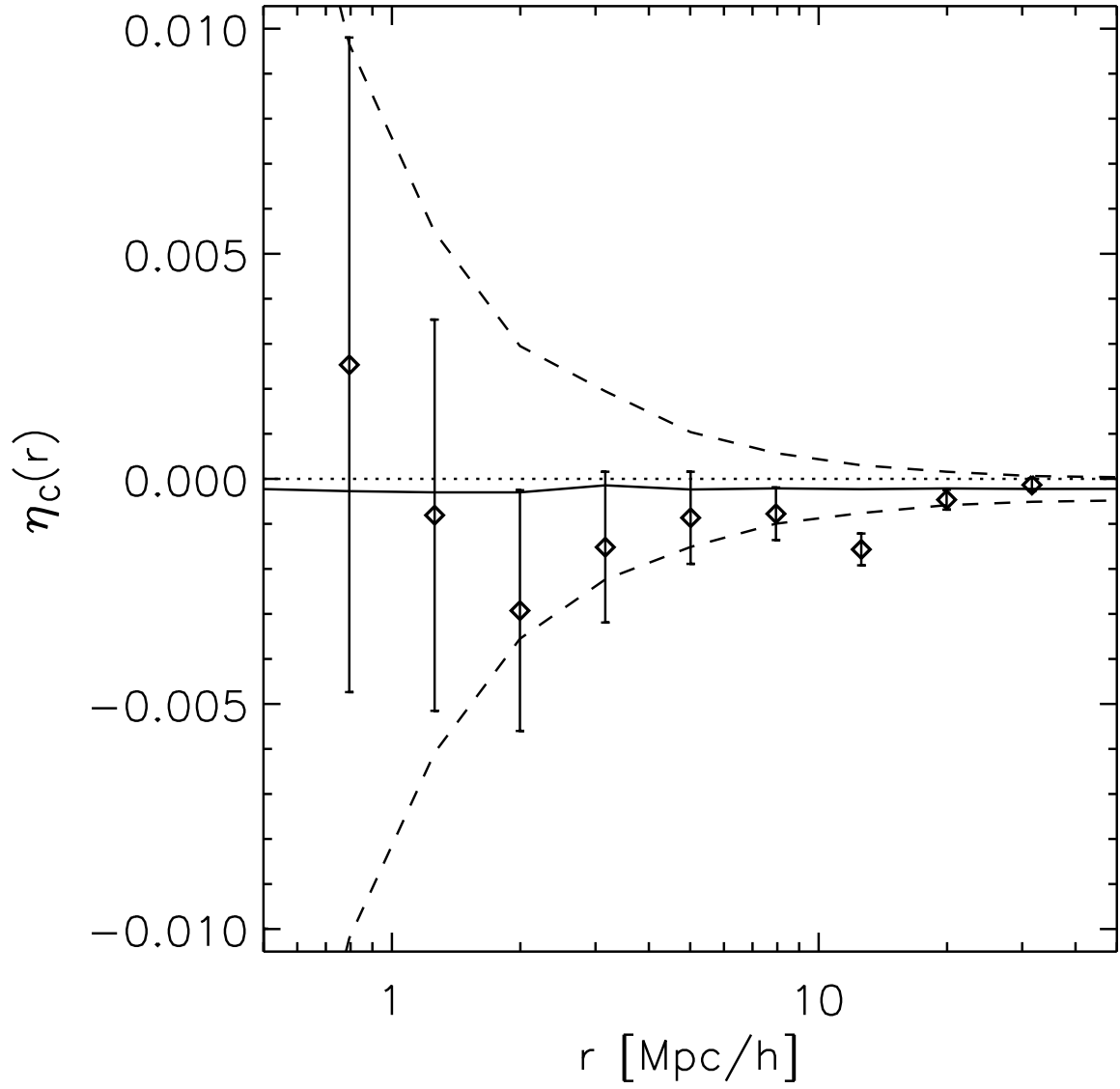}
\caption{Cross-correlations of the unit-spin vectors between the faint and the bright 
Scd galaxies at $0\le z<0.02$.}
\label{fig:spincor_crossmag}
\end{center}
\end{figure}

In \S 3.3, we discuss fully the possible sources of this residual systematic errors involved in the measurement of 
the spin correlations of the fainter galaxies. Before moving on to the discussion on systematics, we also measure 
the spin cross correlations between the fainter and brighter samples, the result of which is plotted in 
Figure \ref{fig:spincor_crossmag}.  As can be seen, a strong anti-correlation of the spin axes as significant as 
$3\sigma_{b}$ is found at $r\approx 10\,h^{-1}$Mpc. In other words, the spin axes of the fainter galaxies 
tend to be orthogonal to that of the brighter galaxies separated by the distance of 
$\sim 10\,h^{-1}$Mpc.  Given the numerical and observational results from the previous works 
\citep{bai-etal05, BS05,jones-etal10,hah-etal10}, we explain this observational result as follows: 
The spin axes of the fainter galaxies are aligned with the intermediate principal axes of 
the local tidal fields as predicted by the linear tidal torque theory. 
Whereas for the brighter galaxies which are usually located in cosmic filaments, their  
spin axes tend to be aligned with the longest axes of the local filaments, i.e., the minor 
principal axes of the local tidal fields \citep{jones-etal10}, as revealed by the recent numerical 
results from hydrodynamic simulations \citep{hah-etal10}. In consequence, the two subsamples have 
strong anti spin-correlations between each other. The separation distance of $10\,h^{-1}$Mpc 
at which the strong anti-correlations are detected should be the typical separation between the 
faint isolated and bright wall galaxies. 

\subsection{Discussion on Systematics}

For the accurate measurement of the galaxy spin correlations, we have tried to eliminate systematic errors 
as much as we can by constraining the redshift, morphological type and angular size of the SDSS galaxies. 
The small bootstrap errors shown in the bottom right panel of Figure \ref{fig:spincor_z} suggests that 
no severe systematics be existent in our constrained sample, and thus that our detection of the spin correlations 
be robust. Nevertheless, it has to be admitted that our results may not be completely free from systematics. 

A possible source of residual systematics could be related to the constant value of the intrinsic flatness 
parameter, $p=0.1$ in Equation (\ref{eqn:xi}) that we adopt in accordance with \citet{HG84} to account for 
the finite thickness of the Scd galaxies. Given that this value of $p=0.1$ was obtained by \citet{HG84} from 
a sample of HI 21cm observations and that the shape of the neutral hydrogen distribution does not necessarily 
coincide with the optical shape of a galaxy,  one may suspect that a different value of $p$ should be used for the 
galaxies from the optical SDSS survey.  An ideal way would be to look at the individual images of the optical 
shapes of the SDSS galaxies and to search for the optimal value to $p$ for the selected Scd galaxies. 
This task, however, would be extremely time consuming and thus should be beyond the scope of this paper. 

\begin{figure}
\begin{center}
\plotone{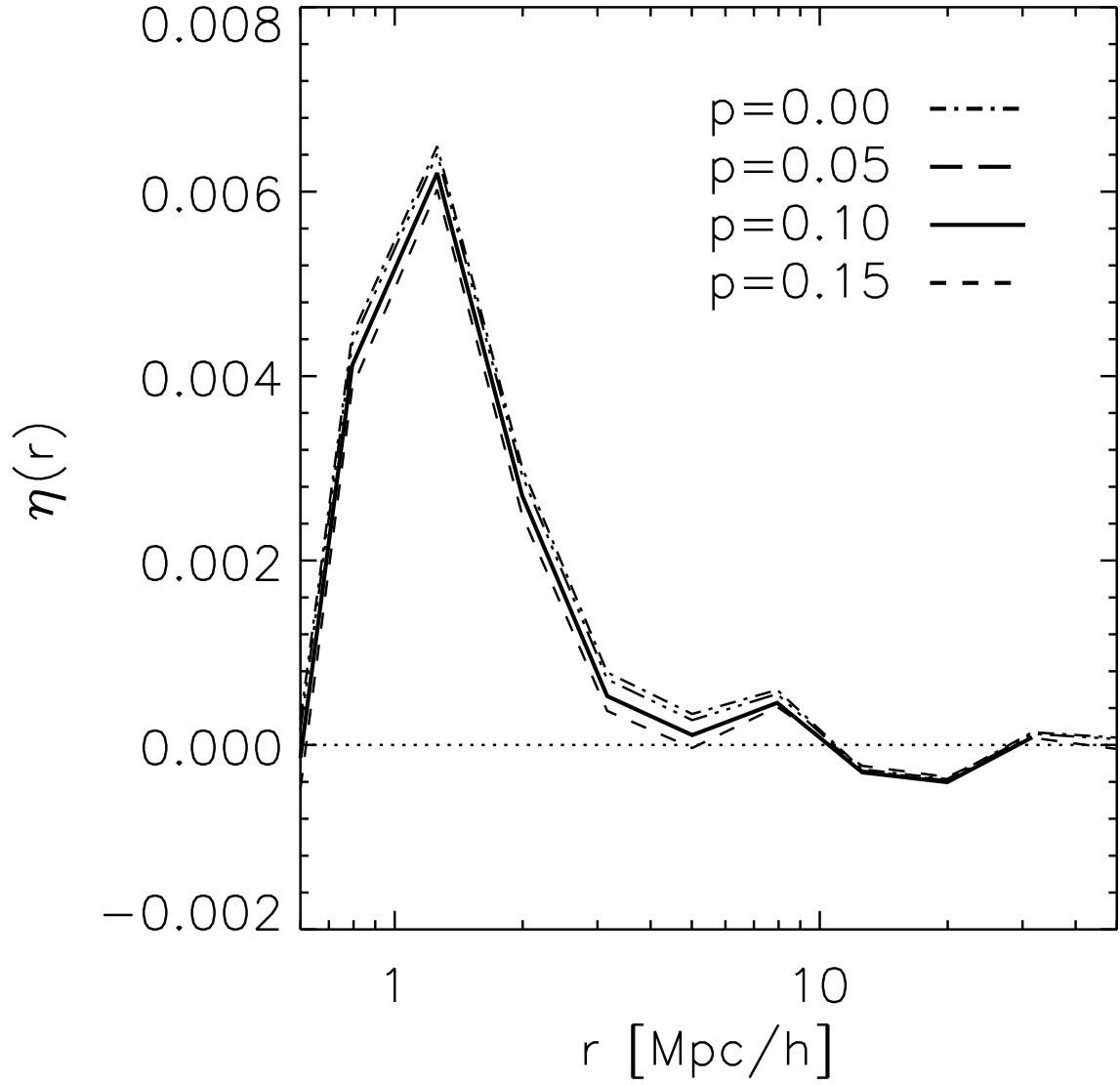}
\caption{Same as the bottom-right panel of Figure \ref{fig:spincor_z} but for four different values of the 
intrinsic flatness parameter, $p$.}
\label{fig:flat}
\end{center}
\end{figure}

Instead, we inspect how the final results would change as we use different values to $p$ in Equation 
(\ref{eqn:xi}). \citet{HG84} explained that unless a galaxy has a well developed bulge like the early type spirals, 
the intrinsic flatness parameter must have the value in range of $0.1\le p\le 0.15$. Since we select only Scd 
galaxies which optical shapes must have the thinnest bulges, we expect that the value of $p$ would not differ 
significantly from $0.1$.  Figure \ref{fig:flat} plots the same as Figure \ref{fig:spincor_z} but for four different 
cases of the intrinsic flatness parameter ($p=0.0\,0.05\, 0.1$ and $0.15$). As one can see, the final results 
change less than $10 \%$ as the value of $p$ changes from $0.$ to $0.15$, which justify our choice of $p=0.1$. 

Another possible source of residual systematics is the limitation of the circular disk approximation. 
The morphology class of Scd from the SDSS catalog includes the irregular galaxies \citep{HC-etal10}, for which 
the circular thin disk approximation would be invalid.  But, through private communication with M.Huertas-Company, 
we have learned that the fraction of the included irregular galaxies must be small since the irregular galaxies are 
usually fainter while the targets of the spectroscopic SDSS  sample are brighter.  
The relatively larger bootstrap mean values of $\eta(r)$ for the fainter samples shown in the top panel 
of Figure \ref{fig:spincor_mag} are most likely to be caused by the inaccurate measurement of the spin axes 
of the irregular galaxies that are included in the subsample of the fainter galaxies ($M_{r}>-16.67$).  
An ideal way should be to look at the individual images of the fainter galaxies and sort out the irregular galaxies 
from the fainter sample. However, it would be also extremely time consuming and thus should be beyond the 
scope of this paper. Here, we just state explicitly that the residual systematics could cause spurious signals 
of correlations for the fainter Scd galaxy sample, which is most likely to be due to the limitation of the 
circular thin disk approximation and that a better algorithm for the measurement of the spin axes of the 
irregular galaxies would be desirable. 

\section{COMPARISON WITH THE THEORETICAL MODEL}

\subsection{A Brief Review of the Analytic Model}

According to the linear tidal torque theory, the angular momentum vector of a proto-galactic 
halo ${\bf L}=(L_{i})$ is expressed in terms of the proto-galaxy inertia momentum tensor 
${\bf I}=(I_{ij})$ and the local tidal shear tensor ${\bf T}=(T_{ij})$ 
\citep{dor70,whi84,CT96,LP00,LP01,cri-etal01,LP02,por-etal02,LE07,LP08}
\begin{equation}
\label{eqn:ang}
L_{i} \propto \epsilon_{ijk}T_{jl}I_{lk},
\end{equation}
where $\epsilon_{ijk}$ is the fully anti-symmetric tensor.  Equations (\ref{eqn:ang}) implies that the spin axes 
of the proto-galaxies (i.e., the direction of the proto-galaxy angular momentum vector) are not random but 
correlated with the intermediate principal axes of the local tidal field tensors \citep{CT96,LP00}. 
Assuming that the initially generated correlations between ${\bf L}$ and ${\bf T}$ have been retained to some 
non-negligible degree till present epoch, \citet{LP00} have proposed the following formula to quantify the 
strength of the correlations:
\begin{equation}
\label{eqn:stc}
\langle\hat{L}_{i}\hat{L}_{j}\rangle = \frac{1+a}{3}\delta_{ij} -
a\hat{T}_{ik}\hat{T}_{kj},
\end{equation}
where $\hat{\bf L}$ is the unit spin vector of a given galaxy, $\hat{\bf T}$ is the unit traceless tidal shear tensor 
smoothed on the galactic scale, and $a$ is a free parameter to quantify the strength of the correlations between 
$\hat{\bf L}$ and $\hat{\bf T}$.

A direct application of Equation (\ref{eqn:stc}) to the spatial correlations between the spin axes of neighbor galaxies 
have yielded the following formula for the galaxy spin-spin correlations \citep{pen-etal00,LP01,cri-etal01}
\begin{equation}
\label{eqn:1st}
\eta(r) \approx \frac{a^2}{6}\frac{\xi^{2}(r;R)}{\xi^{2}(0;R)},
\end{equation}
where $\xi(r)$ is the two-point correlations of the linear density field smoothed on the galactic scale 
($R\approx 1\,h^{-1}$Mpc) \citep[e.g.,see][]{BBKS}. According to Equation (\ref{eqn:1st}), the spatial correlations 
between the galaxy spin axes would follow a quadratic scaling of the density correlation, decreasing rapidly 
to zero as $r$ increases.  We fit the observations results obtained in \S 3.1 and 3.2 to Equation (\ref{eqn:1st}) 
to determine the best-fit correlation parameter, $a$ and examine whether or not the observed spin correlations
follow the quadratic scaling with the linear density correlation, $\xi(r)$.

\subsection{Observed Scaling of the Galaxy Spin-Spin Correlations}

We minimize the following generalized $\chi^{2}$ to find the bestfit-value of $a$ \citep{hartlap-etal07}.
\begin{equation}
\label{eqn:chi2}
\chi^{2}\equiv \frac{1}{(N_{bin}-1)}[\eta_{i}-\eta(r_{i})]C^{-1}_{ij} [\eta_{j}-\eta(r_{j})],
\end{equation}
where $C^{-1}_{ij}$ is the inverse of $C_{ij}$ that is defined in Equation (\ref{eqn:cov}), $\eta(r_{i})$ represents the 
theoretical value calculated at the $i$-th distance bin, $\eta_{i}$ is the observed correlation at the same distance bin 
(with the bootstrap means subtracted) and $N_{bin}$ is the number of radial bins. Note that since there is only one 
parameter, the degree of freedom for $\chi^{2}$ is given as $N_{bin}-1$. 
Here, the covariance matrix $C_{ij}$ is computed using the bootstrap resamples as 
defined in \S 2.2 and the fitting is done for the distance range of $0.5\le r/(h^{-1}{\rm Mpc})\le 50$ 
since at larger distances the numerical flukes tend to contaminate the fitting.
If this {\it reduced} $\chi^{2}$ have a value close to unity, the theoretical model,  
Equation (\ref{eqn:1st}), would be regarded as a good fit to the observational result.
 
\begin{figure}
\begin{center}
\plotone{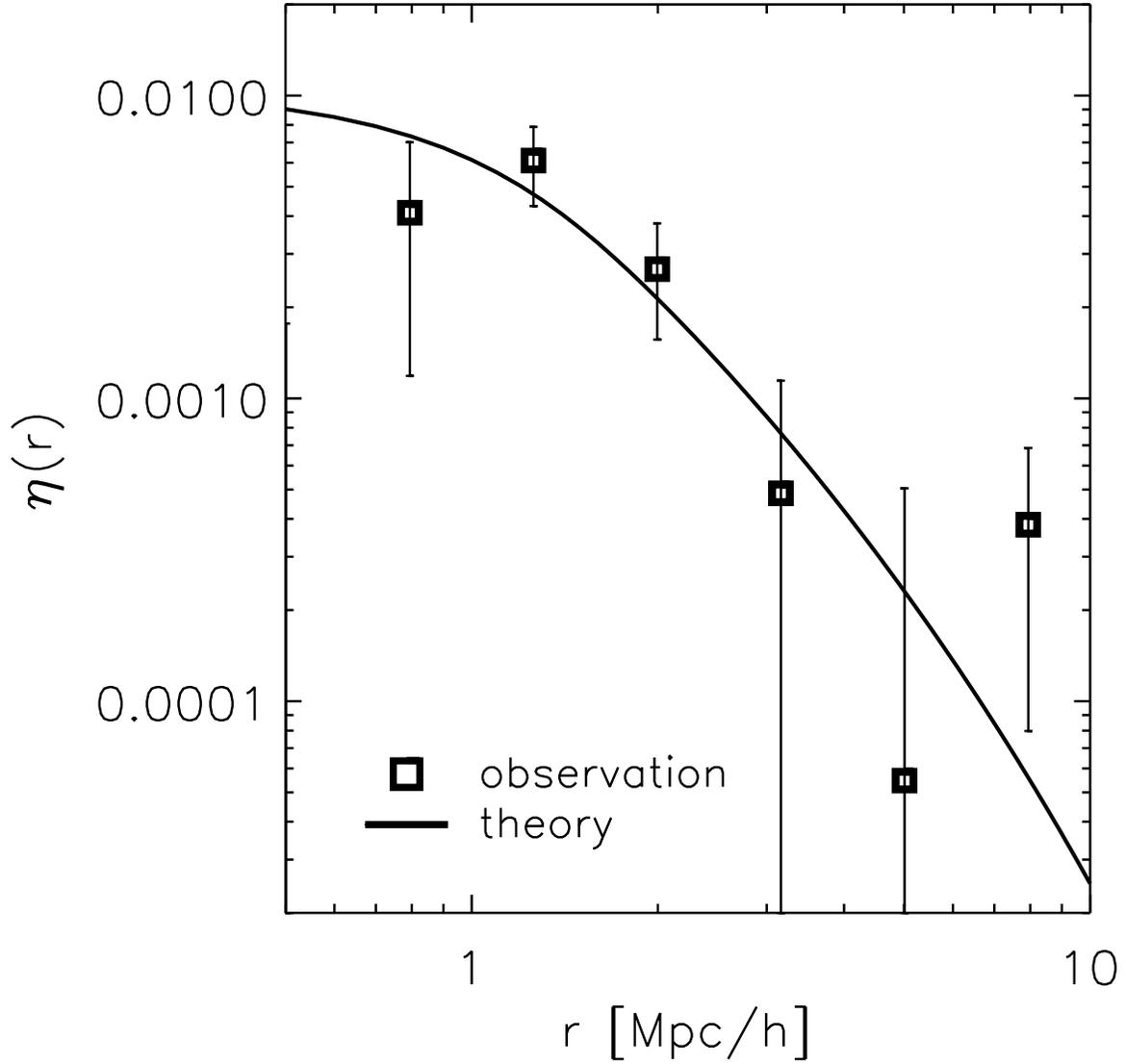}
\caption{Comparison of the spin-correlations of the selected Scd galaxies (square dots) with the theoretical model 
with the best-fit parameter (solid line).  The data points (square dots) are obtained by subtracting the bootstrap 
average from the observed spin-correlation values. The errors represents 1$\sigma$ bootstrap scatter. }
\label{fig:spincor_theory}
\end{center}
\end{figure}

Figure \ref{fig:spincor_theory} plots the best-fit model as solid line and compares it with the observational 
results (square dots). The best-fit value of $a$ is found to be $0.25\pm 0.04$ and the corresponding value of 
$\chi^{2}$ is found to be $0.83$ (see Table \ref{tab:parameter}). Here, the error associated with the measurement 
of $a$ is determined as \citep{hartlap-etal07}
\begin{equation} 
\label{eqn:sigma_a}
\sigma^{2}_{a} = \left(\frac{d^{2}\chi^{2}}{da^{2}}\right)^{-1}
\end{equation}
Now that the reduced $\chi^{2}$ is found to be close to unity and the best-fit value of $a$ is found to be 
$6\sigma_{a}$ higher than zero,  it can be said that the observed spin correlations are indeed tidally induced, 
having quadratic scaling with the linear density correlations, as predicted by the linear tidal torque theory. 

We note here that both of the best-fit value of $a$ are larger than that  determined by 
\citet{LP08} through fitting of the numerical data from the Millennium Run simulations (see Table 3 in 
Lee \& Pen 2008). There are two important implications of this result. First of all, the spin orientations of the 
luminous galaxies may not be perfectly aligned with those of their dark halos, which is consistent with the 
previous numerical results  \citep[e.g.,][]{bai-etal05,BS05,hah-etal10}. Second of all, the luminous 
galaxies have retained better the initial memory of the tidally induced intrinsic spin-spin 
correlations than their dark halos. This phenomenon may be explained as follows. 
The luminous galaxies are located in the inner parts of the galactic halos while the dark matter 
particles are stretched to the outer parts of the dark halos \citep[e.g.,][]{BS05,hah-etal10}. 
Therefore, the spin orientations of the luminous galaxies may be less vulnerable than those 
of their dark counterparts to the destructive nonlinear effects from the surroundings which 
would break the tidally induced intrinsic correlations.  

\begin{figure}
\begin{center}
\plotone{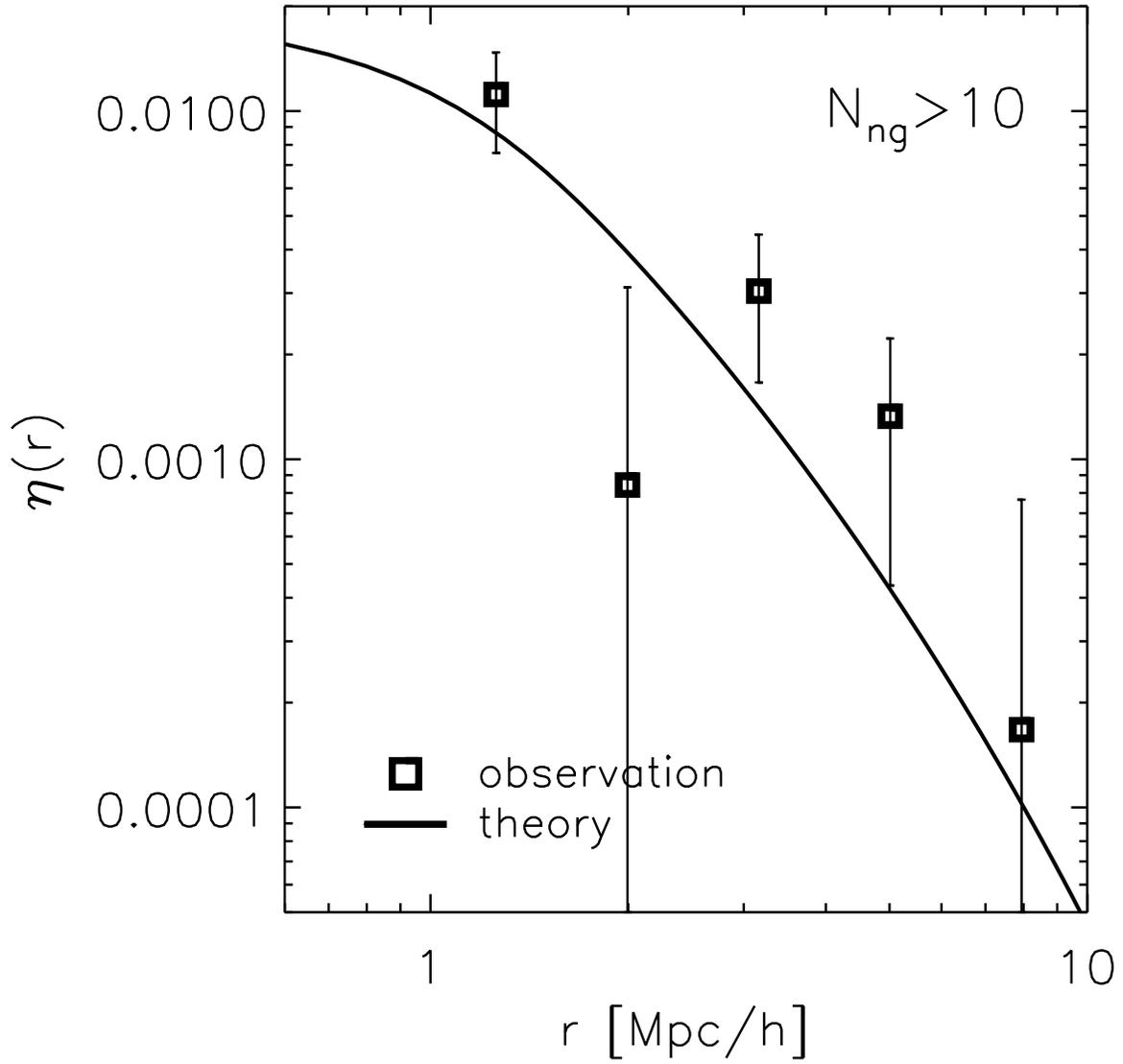}
\caption{Same as Figure \ref{fig:spincor_theory} but only for those Scd galaxies with 
$N_{ng}>10$.}
\label{fig:spincor_hden_theory}
\end{center}
\end{figure}

\begin{figure}
\begin{center}
\plotone{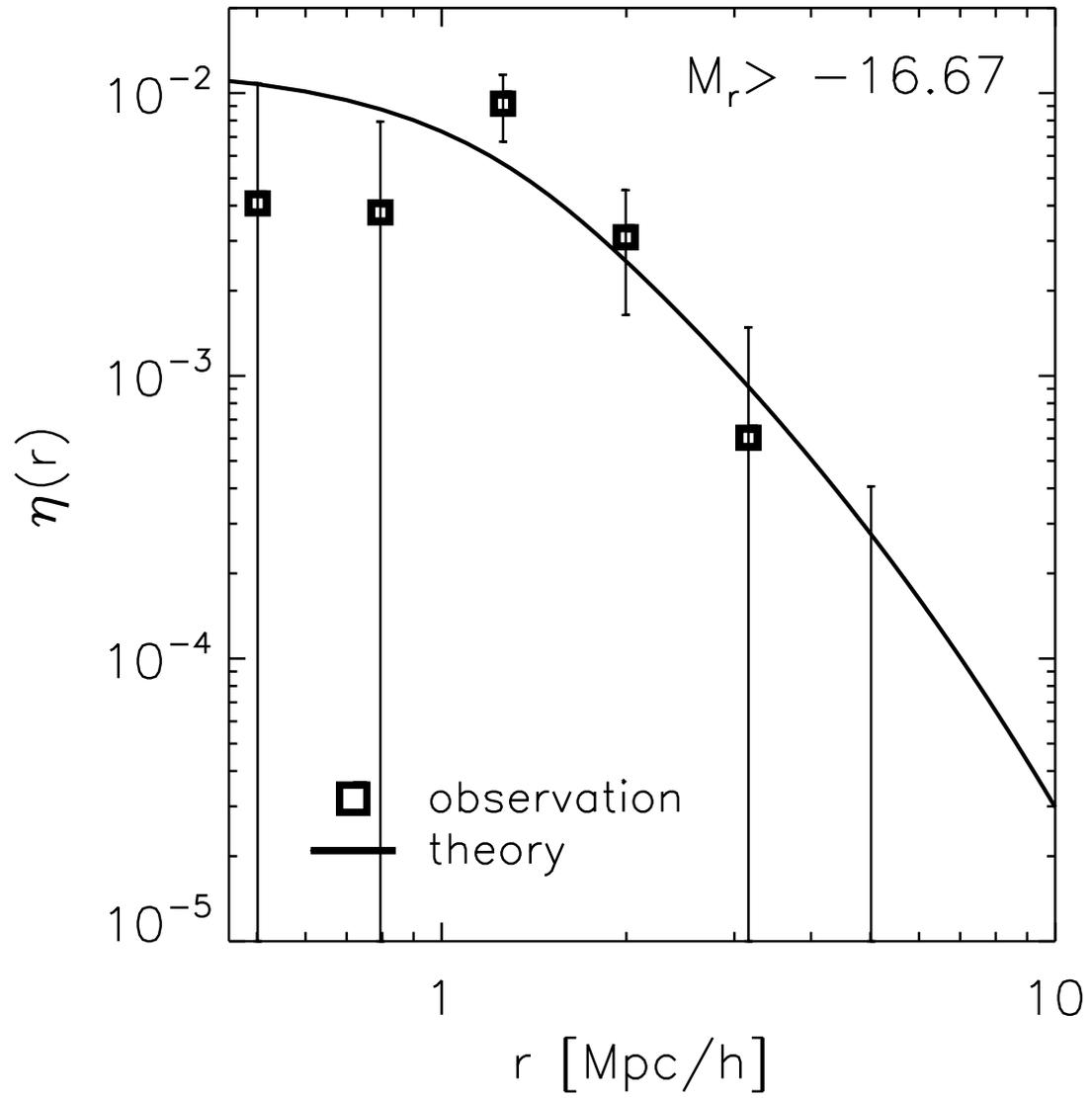}
\caption{Same as Figure \ref{fig:spincor_theory} but only for those Scd galaxies with 
$M_{r}> -16.67$.}
\label{fig:spincor_dim_theory}
\end{center}
\end{figure}

The fitting results for the subsamples with $N_{ng}>10$ and $M_{r}>-16.67$ are also plotted in 
Figures \ref{fig:spincor_hden_theory} and \ref{fig:spincor_dim_theory}, respectively. The best-fit values 
of $a$ and the minimum value of $\chi^{2}$ corresponding to these two cases are also listed in the 
second and third rows of Table \ref{tab:parameter}. As one can see, for the case of $N_{ng}>10$, 
the best-fit value  of $a$ has the largest value, approximately $0.34$.  As mentioned in \S 3.2, 
it may be due to the stronger tidal forces from the surrounding. 

\section{SUMMARY AND CONCLUSIONS}

Selecting those large nearby late-type spiral galaxies which have angular sizes larger than $7.92$ arcsec, 
low redshifts ($0\le z\le 0.02$), and morphological type of Scd from the SDSS DR7, and we have measured 
the spatial correlations between their spin axes. The summary of our results is the following.
\begin{itemize}
\item
A clear signal of the intrinsic correlation as strong as $3.4\sigma_{b}$  and $2.4\sigma_{b}$ 
(where $\sigma_{b}$ is the bootstrap error) is detected at $r\approx 1.25\,h^{-1}$Mpc and 
$r\approx 2.0\,h^{-1}$Mpc, respectively. The cross-correlations between the two radial bins 
are found to be insignificant.
\item
The galaxies having more than ten neighbors within separation of $2\,h^{-1}$Mpc exhibit
higher intrinsic spin correlations (at $3.1\sigma_{b}$ level) than those having ten or less neighbors 
within the same separation distance.  This local density dependence of the galaxy spin correlations 
may be due to the stronger tidal forces from the denser environments. 
\item
The observed spin correlation functions of the selected Scd galaxies follow the quadractic scaling with 
the linear density correlations, which is consistent with the prediction of the linear tidal torque theory.  
\end{itemize}

We have also attempted to measure a luminosity dependence of the galaxy spin correlations 
by splitting the galaxy samples according to their absolute $r$-band magnitudes ($M_{r}$), and 
noted that the correlation function has a peak at $1.25\,h^{-1}$Mpc and $2\,h^{-1}$Mpc for the fainter 
($M_{r}> -16.67$) and brighter sample ($M_{r}\le -16.67$), respectively. However, the two peaks from 
the two magnitude-split samples turn out to correspond to a similar angular separation, which hints at 
significant systematics. To make matters worse, the redshift distributions of the two magnitude-split 
samples are found to be so different that it is impossible with our samples to disentangle a luminosity from 
a redshift dependence of the galaxy spin correlations. Therefore, we conclude that a luminosity dependence 
of the galaxy spin correlations cannot not be measured from our samples.

The behaviors of the detected correlations are found to be consistent with the predictions of the 
tidal torque theory. But, the correlation strength turns out to be greater than predicted by the numerical 
simulations. This result indicates that the spin orientations of the luminous galaxies are not aligned with 
those of the underlying dark halos and that the luminous galaxies retain better the initial memory of  the 
tidally induced spin alignments than their dark counterparts.  It is physically explained as the spin orientations 
of the luminous galaxies may be less vulnerable than those of their dark counterparts to the destructive nonlinear 
effects from the surroundings which tend to break the tidally induced intrinsic correlations.  
It has to be understood fully in the future how and under what circumstances the spin axes 
of the late-type spiral galaxies have retained their original orientations. 

It is also worth mentioning the apparent inconsistency between our results and the previous studies which 
claimed that the two dimensional projected shapes of the blue SDSS galaxies show no correlation at higher 
redshifts, $z\sim 0.1$, \citep[see e.g.,][]{man-etal10}.  Given that we have detected $3.4\sigma$ 
correlation signal at $z\le 0.02$ and that the intrinsic spin correlations of the blue galaxies would be stronger 
at higher redshifts, one may naturally expect that the shape correlations of the blue galaxies at  $z\sim 0.1$ 
would be strong enough to affect the weak lensing signals, which is inconsistent with the previous results. 
We think of several reasons for this inconsistency: First, the two dimensional projected shapes of the blue galaxies 
may not be a good indicator of the tidally induced alignments; Second, there may be considerable uncertainties 
involved in the accurate measurements of the shape correlations of the blue galaxies at $z\sim 0.1$. 
As a final conclusion, our results will shed a new light on the study of the galaxy intrinsic alignments and the 
weak lensing analysis as well.

\acknowledgments

I thank M. Huertas-Company for providing information on the galaxy position angles and 
magnitudes. I also thank an anonymous referee for very useful comments which helped 
me improve significantly the original manuscript. 
This work was supported by the National Research Foundation of Korea (NRF) 
grant funded by the Korea government (MEST, No.2010-0007819). Support for this work was 
also provided by the National Research Foundation of Korea to the Center for Galaxy 
Evolution Research.

Funding for the SDSS and SDSS-II has been provided by the Alfred P. Sloan Foundation, 
the Participating Institutions, the National Science Foundation, the U.S. Department of 
Energy, the National Aeronautics and Space Administration, the Japanese Monbukagakusho, 
the Max Planck Society, and the Higher Education Funding Council for England. The 
SDSS Web Site is http://www.sdss.org/. 

The SDSS is managed by the Astrophysical Research Consortium for the Participating 
Institutions. The Participating Institutions are the American Museum of Natural History, 
Astrophysical Institute Potsdam, University of Basel, University of Cambridge, Case 
Western Reserve University, University of Chicago, Drexel University, Fermi lab, the 
Institute for Advanced Study, the Japan Participation Group, Johns Hopkins University, 
the Joint Institute for Nuclear Astrophysics, the Kavli Institute for Particle 
Astrophysics and Cosmology, the Korean Scientist Group, the Chinese Academy of Sciences 
(LAMOST), Los Alamos National Laboratory, the Max-Planck-Institute for Astronomy (MPIA), 
the Max-Planck-Institute for Astrophysics (MPA), New Mexico State University, Ohio State 
University, University of Pittsburgh, University of Portsmouth, Princeton University, 
the United States Naval Observatory, and the University of Washington.

\clearpage
\clearpage
\clearpage
\clearpage
\clearpage
\clearpage
\clearpage
\clearpage
\clearpage
\clearpage
\clearpage
\clearpage
\clearpage
\clearpage
\clearpage
\clearpage
\begin{deluxetable}{cccc}
\tablewidth{0pt}
\setlength{\tabcolsep}{5mm}
\tablecaption{number of the selected Scd galaxies, mean redshift, mean $r$-band absolute magnitude, 
and mean number of neighbor galaxies within $r_{s}$. }
\tablehead{$N_{\rm Scd}$ & $\bar{z}$ & $\bar{M}_{r}$ & $\bar{N}_{ng}$ } 
\startdata
$4067$ & $0.01$ & $-16.56$ & $27$
\enddata
\label{tab:Scd}
\end{deluxetable}
\clearpage
\clearpage
\begin{deluxetable}{cccc}
\tablewidth{0pt}
\setlength{\tabcolsep}{5mm}
\tablecaption{selection condition, number of the galaxies, median redshift, and the mean redshift }
\tablehead{condition & $N_{g}$ & $z_{m}$ & $\bar{z}$.} 
\startdata
$M_{r}> -16.67$ & $2034$ & $0.007$ & $0.009$\\
$M_{r}\le -16.67$ & $2033$ & $0.015$ & $0.014$\\
\enddata
\label{tab:zdis}
\end{deluxetable}
\clearpage
\begin{deluxetable}{cccc}
\tablewidth{0pt}
\setlength{\tabcolsep}{5mm}
\tablecaption{selection condition, number of the galaxies, best-fit value of $a$, reduced $\chi^{2}$. }
\tablehead{condition & $N_{g}$ & $a$ & $\chi^{2}_{r}$} 
\startdata
--- & $4067$ & $0.25\pm 0.04$ & $0.83$\\
$N_{ng}> 10$ & $1945$ & $0.34\pm 0.07$ & $1.10$\\
$M_{r}> -16.67$ & $2034$ & $0.27\pm 0.04$ & $1.87$\\
\enddata
\label{tab:parameter}
\end{deluxetable}
\end{document}